\documentclass[prd,english,preprintnumbers,amsmath,amssymb,nofootinbib,twocolumn,superscriptaddress]{revtex4-1}

\pdfoutput=1

\usepackage[latin1]{inputenc}
\usepackage{graphicx}
\usepackage{bbm}
\usepackage{amssymb}
\usepackage{amsmath}
\usepackage{tabularx}
\usepackage{slashed}

\usepackage{dsfont}

\def\0#1#2{\frac{#1}{#2}}

\def\s0#1#2{\mbox{\small{$ \frac{#1}{#2} $}}}

\def\CP{{\mathcal P}}
\def\CC{{\mathcal C}}
\def\CT{{\mathcal T}}

\newcommand{\E}{\mathrm{e}}
\newcommand{\I}{\mathrm{i}}
\newcommand{\be}{\begin{eqnarray}}
\newcommand{\ee}{\end{eqnarray}}
\newcommand{\del}{\partial}

\newcommand{\nn}{\nonumber }
\newcommand{\fslash}{\hspace*{-0.2cm}\slash }

\newcommand{\beq}{\begin{equation}}
\newcommand{\eeq}{\end{equation}}
\newcommand{\bea}{\begin{eqnarray}}
\newcommand{\eea}{\end{eqnarray}}

\newcommand{\psib}{\bar{\psi}}

\usepackage{babel}
\makeatother
\begin{document}

\title{Fierz-complete NJL model study:\\ fixed points and phase structure at finite temperature
and density}

\author{Jens Braun}
\affiliation{Institut f\"ur Kernphysik (Theoriezentrum), Technische Universit\"at Darmstadt, 
D-64289 Darmstadt, Germany}
\affiliation{ExtreMe Matter Institute EMMI, GSI, Planckstra{\ss}e 1, D-64291 Darmstadt, Germany}
\author{Marc Leonhardt} 
\affiliation{Institut f\"ur Kernphysik (Theoriezentrum), Technische Universit\"at Darmstadt, 
D-64289 Darmstadt, Germany}
\author{Martin Pospiech}
\affiliation{Institut f\"ur Kernphysik (Theoriezentrum), Technische Universit\"at Darmstadt, 
D-64289 Darmstadt, Germany}

\begin{abstract}
{\it Nambu--Jona-Lasinio}-type models are frequently employed as low-energy models in 
various research fields. With respect to the theory of the strong interaction, this class
of models is indeed often used to analyze the structure of the phase diagram at 
finite temperature and quark chemical potential. The predictions from such models for the 
phase structure at finite quark chemical potential 
are of particular interest as this regime is difficult to access with lattice Monte Carlo approaches.
 In this work, we consider a {\it Fierz}-complete version of a 
{\it Nambu--Jona-Lasinio} model. By studying its renormalization group flow,
we analyze in detail how {\it Fierz}-incomplete approximations affect the predictive power of such model studies. 
In particular, we investigate the curvature of the phase
boundary at small chemical potential, the critical value of the chemical potential above which no spontaneous
symmetry breaking occurs, and the possible interpretation of the underlying dynamics in terms of 
difermion-type degrees of freedom. {We find that the inclusion of four-fermion channels other than the 
conventional scalar-pseudoscalar channel is not only important at large chemical potential but also leaves a significant
{imprint on the dynamics} at small chemical potential as measured by the curvature of the finite-temperature phase boundary.}
\end{abstract}

\maketitle

%
\section{Introduction}
The {\it Nambu--Jona-Lasinio} (NJL) model and its relatives, such as the quark-meson (QM) model, 
play a very prominent role in theoretical physics. Originally, the NJL model has been
introduced as an effective theory to describe spontaneous symmetry breaking in particle physics based on an analogy with 
superconducting materials~\cite{Nambu:1961tp, *Nambu:1961fr}. Since then, it {has frequently been employed}
to study {the phase structure of Quantum Chromodynamics (QCD), i.e. the} theory of the strong interaction,
see, e.g., Refs.~\cite{Klevansky:1992qe,Buballa:2003qv,Fukushima:2011jc,Andersen:2014xxa} for reviews. In particular
at low temperature and large quark chemical potential, NJL-type models have {become an 
important tool to} analyze the low-energy dynamics
of QCD as this regime is at least difficult to access with lattice Monte Carlo techniques.

{NJL/QM-type models indeed provide us with an effective description of 
the chiral low-energy dynamics of QCD, giving us a valuable insight 
into the dynamics underlying the QCD phase diagram.
However, despite the great success of the studies of these models, 
the} phenomenological analysis of the results 
suffers from generic features of {these models as well {as from approximations} underlying
these studies.
For example, NJL-type models in four space-time dimensions 
are defined with an ultraviolet (UV) cutoff~$\Lambda$ 
as they are perturbatively non-renormalizable. In fact, non-perturbative
studies even indicate that they are also not non-perturbatively renormalizable (see, e.g., Refs.~\cite{Braun:2011pp,Braun:2012zq}), in contrast
to three-dimensional versions of this class of models~\cite{Braun:2010tt}.
In case of four space-time dimensions, 
the UV cutoff~$\Lambda$ should therefore be considered as one of the model parameters and also
the regularization scheme belongs to the definition of the model. 
Moreover, we add that often so-called three-dimensional/spatial 
regularization schemes are employed which explicitly break {\it Poincar\'{e}} 
invariance, {potentially leading to} spuriously emerging symmetry breaking patterns in these studies.

The four-quark couplings appearing in a specific ansatz of an NJL-type model are usually considered as fundamental parameters 
and are
fixed such that the correct values of a given set of low-energy observables is reproduced at, e.g., vanishing temperature and quark chemical potential.
Unfortunately, there may exist different parameter sets which reproduce the correct values of a given set of
low-energy observables equally well. Moreover, these model parameters may depend on the external control parameters, such
as the temperature and the quark chemical potential~\cite{Springer:2016cji}.
In any case, even in studies of the conventional NJL/QM model {defined
with only a scalar-pseudoscalar four-quark} interaction channel, other four-quark interaction channels (e.g. a vector channel)
are in general induced due to quantum fluctuations 
but {have often been} ignored in the literature. {In particular at finite chemical potential, 
effective degrees of freedom 
associated with four-quark interaction channels other than the scalar-pseudoscalar channel are expected to become
important or even dominant, see, e.g., Refs.~\cite{Bailin:1983bm,Buballa:2003qv,Alford:2007xm,Anglani:2013gfu} for reviews.
In this work, we moreover demonstrate that such channels may not only play a 
prominent role at large chemical potential but also affect the dynamics at small chemical
potential. In fact, we observe that the inclusion of four-quark channels 
other than the scalar-pseudoscalar channel results in a significantly smaller
curvature of the finite-temperature phase boundary at small chemical potential.
From a field-theoretical point of view, the issue of including more than just
the scalar-pseudoscalar channel is already relevant in the vacuum limit and
is related to the ambiguities associated with {\it Fierz} transformations, i.e. the} fact that 
a given point-like four-quark interaction channel respecting the symmetries of the underlying theory is reducible by means of these transformations. 
As QCD low-energy model studies in general do not take into account a {\it Fierz}-complete basis of four-quark interactions, they are incomplete with
respect to these transformations.
Even worse, mean-field studies of QCD low-energy models show a basic ambiguity related to the possibility to perform {\it Fierz} 
transformations. {Therefore, the} results from these models 
potentially depend on an unphysical parameter which reflects the choice of the mean field and
limits the predictive power of the mean-field approximation~\cite{Jaeckel:2002rm}.

In order to gain a better understanding of how {\it Fierz}-incomplete approximations of QCD low-energy models potentially affect the 
predictions for the phase structure at finite temperature and density, we study a purely fermionic formulation of the NJL model with 
{a single} fermion species
at leading order of the derivative expansion of the effective action. In particular, we take into account
the explicit symmetry breaking arising from the presence of a heat bath and the chemical potential. 

In Sec.~\ref{sec:model}, we discuss our model and aspects of symmetries relevant for our analysis. The renormalization group (RG) fixed-point structure of
the model at zero temperature and density at leading order of the derivative expansion of the effective action is then
discussed in Sec.~\ref{sec:fpsb}, which also includes a discussion of the {relation between the fixed-point structure and spontaneous}
symmetry breaking. In Sec.~\ref{sec:ps}, we finally discuss the phase structure of our model at finite temperature and 
chemical potential and analyze how it is altered when {\it Fierz}-incomplete approximations are considered. In particular, we 
analyze the curvature of the phase boundary at small chemical potential, the critical value of the chemical potential above which no spontaneous
symmetry breaking occurs, and the possible interpretation of the underlying dynamics in terms of 
effective difermion-type degrees of freedom. Our conclusions can be found in Sec.~\ref{sec:conc}.

\section{Model}\label{sec:model}
For studies of the QCD phase structure at finite temperature and density, the most common approximation in terms of NJL/QM-type models
is to consider an action which only consists of a kinetic term for the quarks and a scalar-pseudoscalar four-quark 
interaction channel. The latter is associated with $\sigma$-meson and pion interactions and is usually considered most relevant
for studies of chiral symmetry breaking because of its direct relation to the chiral order parameter. 
In our present work, we shall consider 
a purely fermionic formulation of the NJL model {with a single} fermion species. Clearly, this corresponds to 
a simplification as the number of fermion species is drastically reduced compared to, e.g., QCD with two flavors and three colors.
Still, this simplified model already shares many aspects with QCD in the low-energy limit and
 allows us to analyze {in a more accessible fashion how} neglected four-fermion interaction
channels and the associated issue of {\it Fierz}-incompleteness affect the predictions for the phase structure at finite temperature and density.

In order to relate our present work to conventional QCD low-energy model studies, we 
start our discussion by considering a so-called classical action~$S$ which essentially consists of a kinetic term for the
fermions and a scalar-pseudoscalar four-fermion interaction channel in four Euclidean space-time dimensions:
\be
S[\psib, \psi]&=& \int_0^\beta {\rm d} \tau \int {\rm d}^3 x\, \Big\{  \psib (\I \slashed \del - \I \mu \gamma_0 ) \psi \nn \\ 
&&\qquad\qquad\quad + \frac{1}{2}\bar \lambda_\sigma  \left[ (\psib \psi )^2 - (\psib \gamma_5 \psi)^2 \right]
\Big\}\,.
\label{eq:S}
\ee
Here, $\beta = 1/T$ denotes the inverse temperature~$T$ and~$\mu$ is the chemical potential.
This action is invariant under simple phase transformations,
\be
U_\text V (1) &:& \psib \mapsto  \psib \E^{-\I \alpha}, \enspace \psi\mapsto \E^{ \I \alpha} \psi\,.
\ee
As we do not allow for an explicit fermion mass term, the action is also invariant under 
chiral $U_{\text{A}}(1)$ transformations, i.e. axial phase transformations:
\be
U_\text A (1) &:&  \psib \mapsto \psib  \E^{ \I \gamma_5 \alpha}, \enspace  \psi \mapsto \E^{ \I \gamma_5 \alpha} \psi\,,
\ee
where~$\alpha$ is the ``rotation" angle in both cases. The chiral symmetry is broken spontaneously if a finite vacuum 
expectation value~$\langle\bar{\psi}\psi\rangle$ is generated by quantum fluctuations.
The $U_{\text{V}}(1)$ symmetry is broken spontaneously if, e.g., a difermion
condensate~$\langle \psi^{T}\CC\gamma_5\psi\rangle$ is formed, where $\CC=\I \gamma_2\gamma_0$ is the charge
conjugation operator.

Because of the presence of a heat bath and a chemical potential, {\it Poincar\'e} invariance is explicitly broken and the {\it Euclidean} time 
direction is distinguished. Note also that a finite chemical potential {explicitly breaks the charge} conjugation symmetry~$\CC$. 
However, the rotational invariance among the spatial components  
as well as  
the invariance
with respect to parity transformations~$\CP$ and time reversal transformations $\CT$ remain intact. 

Let us now consider the quantum effective action~$\Gamma$ which is obtained from the path integral by
means of a Legendre transformation. The classical action~$S$ of the theory entering the path integral 
can be viewed as the zeroth-order approximation of the quantum effective
action~$\Gamma$. If we now compute quantum corrections to~$\Gamma$, we immediately observe that four-fermion
interaction channels other than the scalar-pseudoscalar interaction channel are induced, even though they do not appear in the classical
action~$S$ in Eq.~\eqref{eq:S}, see, e.g., Ref.~\cite{Braun:2011pp} for a review. For example, a vector-channel interaction~$\sim (\bar{\psi}\gamma_{\mu}\psi)^2$
may be generated. Once other
four-fermion channels are generated, it is reasonable to expect that these channels also alter dynamically 
the strength of the original scalar-pseudoscalar interaction.
In particular at finite temperature and density, the number of possibly induced interaction channels is even increased because of
the reduced symmetry of the theory.
For our present study of the quantum effective action at leading order (LO) of the derivative expansion, we therefore consider the most general 
ansatz for the effective average action compatible with the symmetries of the {theory:
\be
&& \Gamma_{\rm LO}[\psib, \psi] \nn\\
&& \quad = \int_0 ^\beta {\rm d} \tau  \int {\rm d}^3 x \: \Big\{ 
\psib ( Z^\parallel \I \gamma_0 \del_0 + Z^\perp \I \gamma_i \del_i - Z_\mu \I \mu \gamma_0 ) \psi\nn \\
&&\quad\;\; +\frac{1}{2} Z_{\sigma}\bar\lambda_\sigma (\text{S}-\text{P}) \!-\! \frac{1}{2}Z_{\text{V}}^{\parallel}\bar \lambda ^\parallel _\text V \left( \text{V}_{\parallel} \right)
\!-\!\frac{1}{2}  Z_{\text{V}}^{\perp}\bar\lambda ^\perp _\text V \left( \text{V}_{\perp} \right)\nn\\
&&\quad\quad -\frac{1}{2} Z_{\text{A}}^{\parallel}\bar{\lambda} ^\parallel _\text A \left( \text{A}_{\parallel} \right) \!-\! \frac{1}{2} Z_{\text{A}}^{\perp}
\bar{\lambda} ^\perp _\text A \left( \text{A}_{\perp} \right)
 \!-\! \frac{1}{2} Z_{\text{T}}^{\parallel}\bar \lambda_{\text T}^{\parallel} \left( \text{T}_{\parallel} \right)\!\!\Big\}\,,
\label{eq:gammagen}
\ee
where} $\bar{\lambda}_\sigma$, $\bar{\lambda}^\parallel _\text V$, $\bar{\lambda}^\perp _\text V$, $\bar{\lambda}_{\text{A}}^{\parallel}$, $\bar{\lambda}^\perp _\text A$, 
and~$\bar{\lambda}_{\text T}^{\parallel}$ denote the bare four-fermion couplings which are accompanied by 
their vertex renormalizations~$Z_{\sigma}$, $Z_{\text{V}}^{\parallel}$, $Z_{\text{V}}^{\perp}$, $Z_{\text{A}}^{\parallel}$, 
$Z_{\text{A}}^{\perp}$, and~$Z_{\text{T}}^{\parallel}$, respectively. The various four-fermion interaction channels
are defined as follows:
\be
(\text{S}-\text{P}) &\equiv& (\psib \psi) ^2 - (\psib \gamma_5 \psi)^2\,,\nn\\
 \left(\text{V}_{\parallel}\right) &\equiv& (\psib \gamma_0 \psi)^2\,,\quad
\left(\text{V}_{\perp}\right) \equiv ( \psib \gamma_i \psi)^2\,,\nn\\
 \left(\text{A}_{\parallel}\right) &\equiv& (\psib \gamma_0 \gamma_5 \psi)^2\,,\quad
 \left(\text{A}_{\perp}\right)\equiv ( \psib \gamma_i \gamma_5 \psi)^2\,,\nn\\
 \left(\text{T}_{\parallel}\right) &\equiv& (\psib \sigma_{0 i} \psi)^2 - (\psib \sigma_{0 i} \gamma_5 \psi)^2\,,
\ee
where $\sigma_{\mu\nu}=\frac{\I}{2}[\gamma_{\mu},\gamma_{\nu}]$ and 
summations over $i=1,2,3$ are tacitly assumed.
The renormalization factors associated with the kinetic term are given by~$Z^\parallel$ and $Z^\perp$, respectively. Finally, the 
chemical potential is accompanied by the renormalization factor~$Z_{\mu}$. At~$T=0$, we 
have~$Z_{\mu}^{-1}\!=\!Z^{\parallel}\!=\!Z^{\perp}$ for~$\mu_{\rm r}\!\equiv\! Z_{\mu}\mu\! <\! m_{\rm f}$. Here,~$m_{\rm f}$ is the 
potentially 
generated renormalized (pole) mass of the fermions:~$m_{\rm f}=\bar{m}_{\rm f}/Z^{\perp}$ with~$\bar{m}_{\rm f}$ being 
the bare fermion mass. The relation between~$Z_{\mu}$ and~$Z^{\perp}$ 
is a direct consequence of the so-called {\it Silver-Blaze} property of quantum field theories~\cite{Cohen:2003kd}.

The ansatz~\eqref{eq:gammagen} is overcomplete. By exploiting the {\it Fierz} identities detailed
in App.~\ref{App:FierzIdentities}, we can reduce the overcomplete set of four-fermion interactions 
in Eq.~\eqref{eq:gammagen} to a minimal {\it Fierz}-complete set:%
\be
&& \Gamma_{\rm LO}[\psib, \psi] \nn\\
&& \quad = \int_0 ^\beta {\rm d} \tau  \int {\rm d}^3 x \: \Big\{ 
\psib ( Z^\parallel \I \gamma_0 \del_0 + Z^\perp \I \gamma_i \del_i - Z_\mu\I  \mu \gamma_0 ) \psi\nn \\
&&\quad +\frac{1}{2} Z_{\sigma}\bar\lambda_\sigma (\text{S}-\text{P}) \!-\! \frac{1}{2}Z_{\text{V}}^{\parallel}\bar \lambda ^\parallel _\text V \left( \text{V}_{\parallel} \right)
\!-\!\frac{1}{2} Z_{\text{V}}^{\perp}\bar \lambda ^\perp _\text V \left( \text{V}_{\perp} \right)\Big\}\,.
\label{eq:FierzCompleteAnsatz}
\ee
Any other pointlike four-fermion interaction invariant under the symmetries of our model
is indeed reducible by means of {\it Fierz} transformations. Fermion self-interactions of higher order (e.g. eight fermion interactions) 
may also be induced due to quantum fluctuations 
at leading order of the derivative expansion\footnote{The leading order of the derivative expansion corresponds
to treating the fermion self-interactions in the pointlike limit, see also our discussion below.}
but do not contribute to the RG flow of the four-fermion couplings at this order and are therefore not included in 
our ansatz~\eqref{eq:FierzCompleteAnsatz}, see Ref.~\cite{Braun:2011pp} for a detailed discussion.

In the following we shall study the RG flow of the four-fermion couplings appearing in the effective action~\eqref{eq:FierzCompleteAnsatz}.
This already allows us to gain a valuable insight into the phase structure of our model.

\section{Vacuum fixed-point structure and spontaneous symmetry breaking}\label{sec:fpsb}

Before we actually analyze the fixed-point structure of our model and its phase structure at finite temperature
and chemical potential, we briefly discuss how a study of the quantum effective action~\eqref{eq:FierzCompleteAnsatz}
at leading order of the derivative expansion can give us access to the phase structure of our model at all. A detailed discussion
can be found in, e.g., Ref.~\cite{Braun:2011pp}. 

{The leading order} of the derivative expansion implies that we treat the four-fermion interactions in the 
pointlike limit,\footnote{Note that the anomalous dimensions associated with the wavefunction renormalizations~$Z^{\parallel}$
and~$Z^{\perp}$ vanish at leading order of the derivative expansion, see our discussion below and also App.~\ref{app:RG} for a 
discussion of possible issues with the derivative expansion in the presence of a finite chemical potential.} i.e. 
\be
&&\!\bar{\lambda}_j (\bar{\psi}{\mathcal O}_j\psi)^2  \nn\\ 
&& =\! \lim_{\{p_k\to 0\}} \!
\bar{\psi}_a(p_1)\bar{\psi}_b(p_2)\Gamma^{(4)}_{j,abcd}(p_1,p_2,p_3,p_4)\psi_c(p_3)\psi_d(p_4)\,,\nn
\ee
where~$a, b, c, d$ are spinor indices, ${\bar{\lambda}}=\{\bar{\lambda}_{\sigma},\bar{\lambda}_{\text{V}}^{\parallel},\bar{\lambda}_{\text{V}}^{\perp}\}$, 
and~${\mathcal O} = \{(\text{S}-\text{P}),\left(\text{V}_{\parallel}\right),\left(\text{V}_{\perp}\right)\}$. 

Apparently, the leading order of the derivative expansion does not give us access to the mass spectrum of our model
which is encoded in the momentum structure of the correlation functions. In particular, the formation of
fermion condensates associated with spontaneous symmetry breaking is indicated by singularities in
the four-fermion correlation functions. 
Thus, this order of the derivative expansion does not allow us to study regimes of the theory in which one of its
symmetries is broken spontaneously. However, it can be used to study the symmetric phase of our model, e.g.
the dynamics at high temperature and/or high density where the symmetries are expected to remain intact.
By lowering the temperature at a given value of the chemical potential, we can then determine a
critical temperature~$T_{\rm cr}$ below which the pointlike approximation
breaks down and a {condensate related to a spontaneous breaking of one} of the symmetries may be generated. 
This line of argument has indeed already
been successfully applied to compute the many-flavor phase diagram of gauge theories~\cite{Gies:2005as,Braun:2005uj,Braun:2006jd,Braun:2014wja}.
We add that the phenomenological meaning of~$T_{\rm cr}(\mu)$ obtained from such an analysis is difficult to assess. 
To be more specific, various possible symmetry breaking patterns exist and the breakdown of the pointlike approximation cannot 
unambiguously be related to the spontaneous breakdown of a specific symmetry, even in our
simple model, see Eq.~\eqref{eq:FierzCompleteAnsatz}. For example, the critical temperature may be associated with, e.g., chiral
symmetry breaking or with spontaneous symmetry breaking in the vector channel. We shall discuss {this issue again below
in more detail.}

The breakdown of the pointlike approximation can be indeed used to detect the onset of spontaneous symmetry breaking.
This can be most easily seen by considering a {\it Hubbard-Stratonovich}-transformation~\cite{Hubbard:1959ub, *Stratonovich}.
With the aid of this transformation, composites of two fermions can be treated as auxiliary bosonic degrees of freedom, e.g. $\bar{\psi}\psi \mapsto \sigma$.
On the level of the path integral, the four-fermion interactions of a given theory are then replaced by terms bilinear in the
so introduced auxiliary fields and corresponding 
{\it Yukawa}-type interaction terms between the auxiliary fields and the fermions. Formally, we have
\be
\!\!\!\!\!\!\bar{\lambda}_j (\bar{\psi}{\mathcal O}_j\psi)^2 \mapsto \sum_a\frac{1}{\bar{\lambda}_j}\phi^{(j)}_a\phi^{(j)}_a
+ \sum_{a,b,c}\bar{\psi}_b \bar{h}_{j}\tilde{{\mathcal O}}_{j}^{abc}\phi^{(j)}_a \psi_c\,.
\label{eq:hst}
\ee
Here, the couplings $\bar{h}_j$ denote the various {\it Yukawa} couplings.
The structure of the quantity~$\tilde{{\mathcal O}}_{j}^{abc}$ 
with respect to internal indices may be nontrivial and depends 
on the corresponding four-fermion interaction channel~${\mathcal O}_j$. The same holds for the exact transformation properties of the
possibly multi-component auxiliary field~$\phi^{(j)}_a$. 

Once a {\it Hubbard-Stratonovich} transformation has been performed, the {\it Ginzburg-Landau}-type effective potential for the bosonic fields~$\phi^{(j)}_a$ 
can be computed conveniently, allowing for a straightforward analysis of the ground-state properties of the 
theory under consideration. For example, a nontrivial minimum 
of this potential indicates the spontaneous breakdown of the symmetries associated with those fields which acquire
a finite vacuum expectation value. 

From Eq.~\eqref{eq:hst}, we also deduce that the four-fermion couplings are 
inverse proportional to the mass-like parameters~$m^2_j\sim 1/\bar{\lambda}_j$ 
associated with terms bilinear in the bosonic fields. Recall now that the transition from the symmetric regime
to a regime with spontaneous symmetry breaking is indicated by a qualitative change of the shape 
of the {\it Ginzburg-Landau}-type effective potential as some fields acquire a finite vacuum expectation value. 
In fact, in case of a second-order transition, at least one of the curvatures~$m^2_j$ of the effective potential 
at the origin changes its sign
at the transition point. This is not the case for a first-order transition.
Still, {taking into account all quantum fluctuations, the} {\it Ginzburg-Landau}-type effective potential becomes
convex in any case, implying that the curvature tends to zero at both a first-order {as well as a second-order}
phase transition point.
Thus, a singularity of a pointlike four-fermion
coupling indicates the onset of spontaneous symmetry breaking.
However, it does not allow to resolve the nature of the transition. 

For our RG analysis, it follows that the
observation of a divergence of a four-fermion coupling {at an RG} scale~$k_{\rm cr}$ 
only serves as an indicator for the onset of spontaneous symmetry breaking. Below, we shall use this criterion 
to estimate the phase structure of our model. For a given chemical potential, the above-mentioned critical temperature~$T_{\rm cr}$ is then
given by the temperature at which the divergence occurs at~$k_{\rm cr}\to 0$.
Note that this is not a sufficient criterion for spontaneous {symmetry breaking 
as quantum fluctuations may restore the symmetries
of the theory in the deep infrared (IR) limit, see, e.g., Ref.~\cite{Braun:2011pp} for a detailed discussion.
If the true phase transition is of first order, this criterion at leading order of the derivative expansion may even only point
to the onset of a region of metastability and not to the actual phase transition line.} 
{From a QCD standpoint, this implies that the liquid-gas phase transition, which is expected to be of first order, 
cannot be reliably assessed in the setup underlying our present work but requires to extend the truncation of the effective action.
(Color-)Superconducting ground states can in principle be detected within our present setup, 
if the transition is of second order (following our line of arguments above). Indeed, we show that the scaling behavior of physical
observables associated with a superconducting ground state can be recovered correctly from our analysis of the RG flow
of four-fermion couplings, see also Sec.~\ref{sec:ps}.
Thus, despite the discussed restrictions of our present analysis, it}
already {provides a valuable insight} into the dynamics underlying spontaneous symmetry breaking of a given fermionic theory. 

Instead of using the purely fermionic formulation of our model, one may be tempted to consider the partially bosonized formulation of our 
model right away in order to compute conveniently the {\it Ginzburg-Landau}-type effective potential for the various auxiliary fields, as
indicated above.
However, in contrast to the purely fermionic formulation, in which {\it Fierz} completeness at, e.g., leading order of the derivative expansion 
can be fully preserved
by using a suitable basis of four-fermion interaction channels, 
conventional approximations entering studies of the
partially bosonized formulation may {easily induce a so-called {\it Fierz} ambiguity}. Most prominently, mean-field approximations
are known to show a basic ambiguity related to the possibility to perform {\it Fierz} 
transformations~\cite{Jaeckel:2002rm}. {Therefore, results} from this approximation potentially depend 
on an unphysical parameter which is associated with 
the choice of the mean field and limits the predictive power of this approximation.
However, it has been shown~\cite{Jaeckel:2002rm} that the {use of so-called 
dynamical bosonization techniques~\cite{Gies:2001nw,Gies:2002hq,Pawlowski:2005xe,Braun:2008pi,Floerchinger:2009uf, %
Floerchinger:2010da,Braun:2014ata,Mitter:2014wpa} allow to}
resolve this issue,
see also Ref.~\cite{Gies:2006wv} for an introduction to dynamical bosonization in RG flows.
As this is beyond the scope of the present work, we focus exclusively on the purely fermionic formulation of our model.

{With these prerequisites, 
let} us now begin with a discussion of the RG flow of our model in the limit~$T\to 0$ and~$\mu\to 0$.
For the computation of the RG flows of the various four-fermion couplings and wavefunction renormalizations at leading order 
in the derivative expansion, we employ an RG equation for the quantum effective action, the {\it Wetterich} equation~\cite{Wetterich:1992yh}. 
The effective action~$\Gamma$ then depends on the RG scale~$k$ (IR cutoff scale) which determines {the RG ``time"~$t=\ln(k/\Lambda)$}
with~$\Lambda$ being a UV cutoff scale, see App.~\ref{app:RG} and also 
Ref.~\cite{Braun:2011pp} for an introduction to the computation
of RG flows of fermion self-interactions. 

To regularize the loop integrals, we {employ a four-dimensional regularization} 
scheme which is parametrized in our RG approach in form of an exponential
regulator function, see App.~\ref{app:RG} for details. In 
the limit~$T\to 0$ and~$\mu\to 0$, our regularization scheme becomes covariant which is of
great importance. 
To be more specific, so-called spatial regularization schemes, which leave the temporal direction unaffected and
are often used in, e.g., model studies, introduce an explicit breaking of {\it Poincar\'{e}} invariance which is present 
even in the limit~$T\to 0$ and~$\mu\to 0$. This leads to a contamination of the results in this limit. 
This is particularly severe since this 
limit is in general also 
used to fix the model parameters. In principle, one may solve this problem by taking care of the symmetry violating terms with the aid of
corresponding ``Ward identities" or, equivalently, one can add appropriate counter-terms such that the theory remains
{\it Poincar\'{e}}-invariant in the limit~$T\to 0$ and~$\mu\to 0$, see Ref.~\cite{Braun:2009si}. However, 
we have observed that
the predictions for the phase structure are significantly spoilt when a spatial regularization scheme is used without properly
taking care of the associated symmetry-violating terms in the limit~$T\to 0$ and~$\mu\to 0$ (see App.~\ref{app:RG} for details). Therefore we 
have chosen a scheme which respects {\it Poincar\'{e}} invariance in the limit~$T\to 0$ and~$\mu\to 0$. 

With respect to RG studies, we add that, apart from the 
fact that spatial regularization schemes explicitly break {\it Poincar\'{e}} invariance, they lack locality in the temporal direction, i.e.
all time-like momenta are taken into account at any RG scale~$k$ whereas spatial momenta are restricted
to small momentum shells around the scale~$k\simeq |\vec{p}^{\,}|$. Loosely speaking, fluctuation effects are therefore washed out
by the use of this class of regularization schemes and the construction of meaningful expansion schemes of the effective action is complicated
due to this lack of locality. 
\begin{figure*}[t]
\begin{center}
  \includegraphics[trim={.25cm 0 0 0},clip,width=0.35\linewidth]{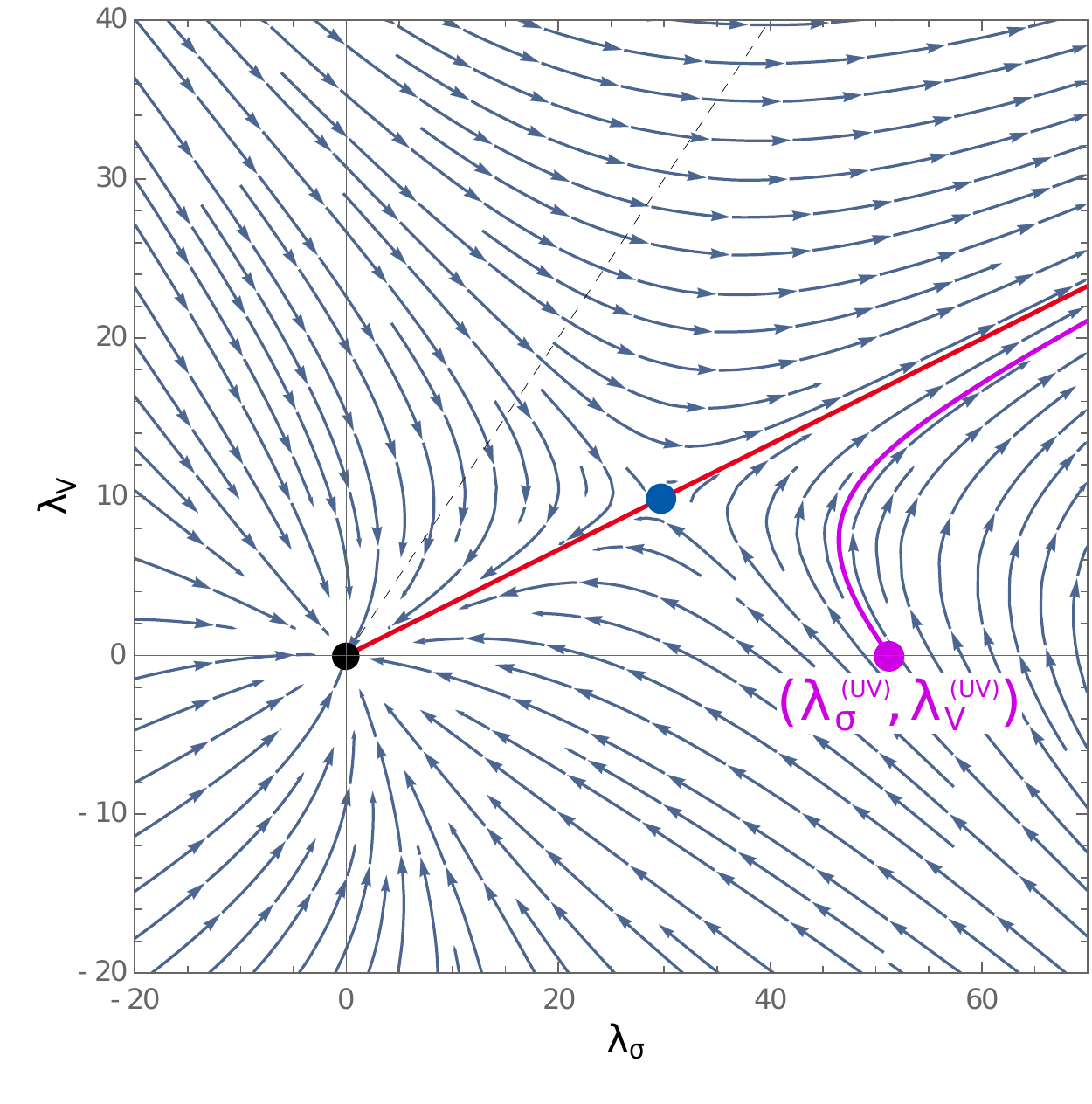}
 \hfill
  \includegraphics[trim={.075cm 0 0 0},clip,width=0.62\linewidth]{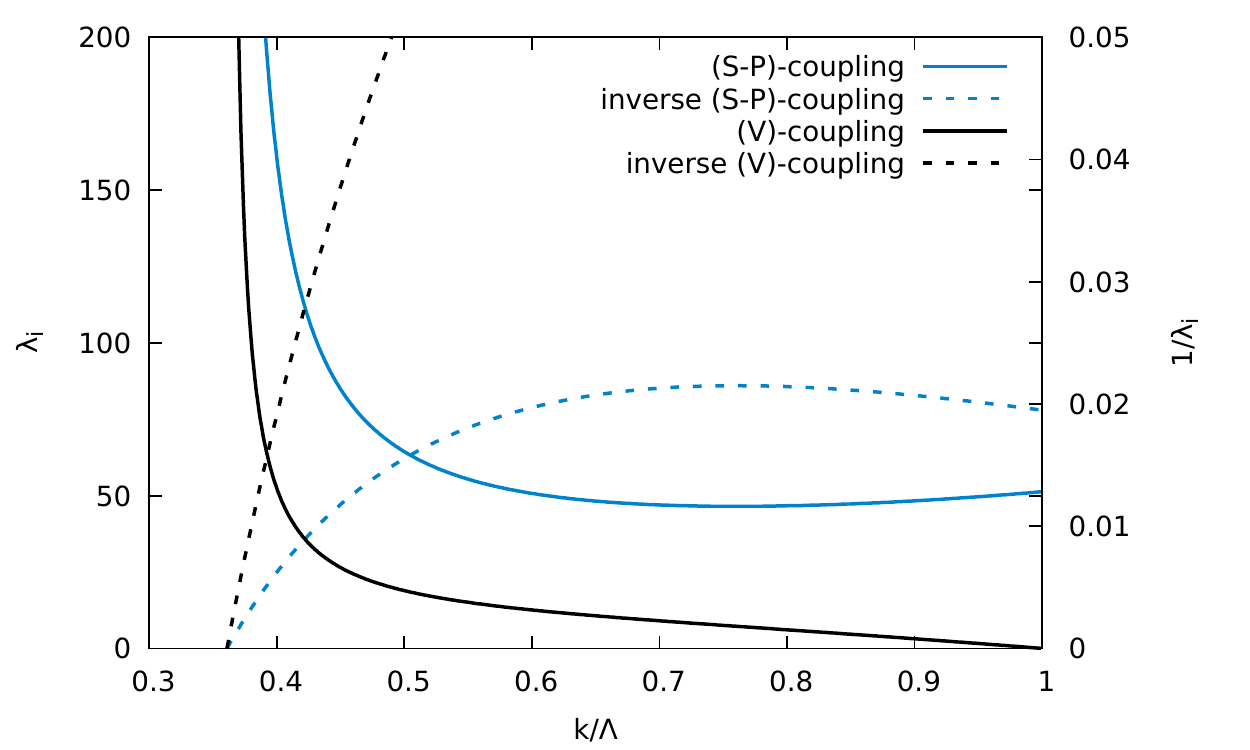}
\end{center}
\caption{Left panel: RG flow at zero temperature
and  chemical potential in the plane spanned by the scalar-pseudoscalar coupling~$\lambda_{\sigma}$ and the vector-channel coupling~$\lambda_{\text{V}}$.
The black dot depicts the Gau\ss ian fixed-point whereas the blue dot depicts one of the two non-Gau\ss ian fixed-points. 
The {pink line} represents an example of an 
RG trajectory. This particular trajectory describing four-fermion couplings diverging at a finite scale~$k_{\rm cr}$  
approaches a separatrix (red line) for~$k\to k_{\rm cr}$.  
The dominance of the scalar-pseudoscalar interaction channel 
is illustrated by the position of this separatrix relative to the bisectrix (dashed back line). Right panel:  
RG scale dependence of the four-fermion couplings~$\lambda_{\sigma}$ and~$\lambda_{\text{V}}$ 
corresponding to the RG trajectory depicted by {the pink line} in the left panel.
The inverse of these two four-fermion couplings
associated with the mass-like parameters~$m_i^2\sim 1/\lambda_i$ of 
terms bilinear in the auxiliary fields in a {\it Ginzburg-Landau}-type effective potential is shown by the dashed lines.}
\label{fig:flowt0}
\end{figure*}

Let us now analyze the fixed-point structure of our model in leading order of the derivative expansion at
zero temperature and chemical potential. 
In this {\it Poincar\'{e}}-invariant limit, the couplings $\lambda_\text V ^\parallel$ and $\lambda_\text V ^\perp$ can be 
identified,~$\lambda_\text V ^\parallel = \lambda_\text V ^\perp =\lambda_{\text{V}}$, provided the two couplings assume
the same value at the initial RG scale~$k=\Lambda$. 
The $\beta$ functions then simplify to\footnote{Note that, for 
a spatial regularization scheme, we find~$\lambda_{\text V} ^\parallel \neq \lambda_{\text V}^{{\perp}}$ even
for~$T=\mu=0$ since such a scheme explicitly breaks {\it Poincar\'{e}} invariance.} 
\be
\del_t \lambda_\sigma &=& \beta _{\lambda _\sigma} = 2 \lambda_\sigma- 8 v_4 \left(  \lambda_\sigma ^2 + 4 \lambda_\sigma  \lambda_\text{V} + 3\lambda_\text{V} ^2 \right),
\label{eq:dtlst0}
\\
\del_t \lambda_\text V &=& \beta _{\lambda _\text{V}} = 2 \lambda _\text{V} 
-4 v_4 \left( \lambda_\sigma + \lambda_\text{V}   \right)^2\,,
\label{eq:dtlvt0}
\ee
where~$v_4=1/(32\pi^2)$ and the dimensionless renormalized four-fermion couplings
are defined as~$\lambda_i=Z_i k^2\bar{\lambda}_i/(Z^{\perp})^2$ 
with~$\bar{\lambda}=\{\bar{\lambda}_{\sigma},\bar{\lambda}_{\text{V}}^{\parallel},\bar{\lambda}_{\text{V}}^{\perp}\}$ 
and~$Z=\{ Z_{\sigma}, Z_{\text{V}}^{\parallel}, Z_{\text{V}}^{\perp}\}$.
Up to regularization-scheme dependent factors, this set of equations agrees with the one found in previous 
vacuum studies of this model~\cite{Jaeckel:2002rm,Braun:2011pp}.
The wavefunction renormalizations, which we have set to~$Z^{\parallel}=Z^{\perp}=1$ at the initial RG scale, 
remain unchanged in the RG flow at this order of the derivative expansion, i.e.~$\partial_t Z^{\parallel}=\partial_t Z^{\perp}=0$.

Before we now discuss the dynamics of the {\it Fierz}-complete system, it is instructive to consider a one-channel approximation. To this end, we 
set~$\lambda_{\text{V}}=0$ in Eq.~\eqref{eq:dtlst0} and drop the flow equation for the vector-channel coupling~$\lambda_{\text{V}}$. Thus,
we are left with the following flow equation for the scalar-pseudoscalar coupling:
\be
\del_t \lambda_\sigma &=& \beta_{\lambda_\sigma} = 2 \lambda_\sigma - 8 v_4 \lambda_\sigma^2\,,
\label{eq:lsca}
\ee
which has a non-Gau\ss ian fixed-point at $\lambda_\sigma^\ast = 8\pi^2$. The solution for~$\lambda_{\sigma}$ reads
\be
\lambda_{\sigma}(k)=\frac{\lambda_{\sigma}^{\text{(UV)}}}{ \left(\frac{\Lambda}{k}\right)^{\Theta}\left(1- \frac{\lambda^{\text{(UV)}}_{\sigma}}{\lambda^{\ast}_{\sigma}}\right) + \frac{\lambda^{\text{(UV)}}_{\sigma}}{\lambda^{\ast}_{\sigma}}}\,.
\label{eq:sol1c}
\ee
Here,~$\lambda^{\text{(UV)}}_{\sigma}$ is the initial condition for the coupling~$\lambda_{\sigma}$ at the UV scale~$\Lambda$
and~$\Theta$ denotes the critical exponent which governs the scaling behavior of physical observables 
close to the ``quantum critical point"~$\lambda_{\sigma}^{\ast}$:
\be
\Theta:=-\frac{\partial \beta_{\sigma}}{\partial \lambda_{\sigma}}\bigg|_{\lambda_{\sigma}^{\ast}}=2\,.
\ee
As~$\Theta >0$, the fixed point~$\lambda_{\sigma}^{\ast}$ is IR repulsive. Indeed,
we readily observe from the solution~\eqref{eq:sol1c} that~$\lambda_{\sigma}$ is repelled by the fixed point. Moreover,~$\lambda_{\sigma}$ 
diverges at a finite RG scale~$k_{\rm cr}$, if~$\lambda^{\text{(UV)}}_{\sigma}$ is chosen to be greater
than the fixed-point value~$\lambda^{\ast}_{\sigma}$, $\lambda^{\text{(UV)}}_{\sigma} > \lambda^{\ast}_{\sigma}$. 
Thus, by varying the {initial condition~$\lambda_{\sigma}^{\text{(UV)}}$,
we can} induce a ``quantum phase transition", i.e. a phase transition in the vacuum limit, from a symmetric phase to 
a phase governed by spontaneous symmetry breaking.

Following our discussion above,
the appearance of a divergence for~$\lambda^{\text{(UV)}}_{\sigma} > \lambda^{\ast}_{\sigma}$
signals the onset of spontaneous symmetry breaking. The associated critical scale~$k_{\rm cr}$ is
given by
\be
k_\text{cr} = \Lambda\left( \frac{ \lambda^{\text{(UV)}}_{\sigma} - \lambda^{\ast}_{\sigma}}{\lambda^{\text{(UV)}}_{\sigma}} \right)^{\frac{1}{\Theta}}
\theta(\lambda^{\text{(UV)}}_{\sigma} - \lambda^{\ast}_{\sigma})
\,.
\label{eq:plaw}
\ee
We emphasize that 
this quantity sets the scale for all low-energy quantities~${\mathcal Q}$ with mass dimension~$d_{{\mathcal Q}}$ 
in our model,~${\mathcal Q}\sim k_{\rm cr}^{d_{{\mathcal Q}}}$.  

Let us now turn to the discussion of the {\it Fierz}-complete system by studying
the flow equations~\eqref{eq:dtlst0} and~\eqref{eq:dtlvt0}.
This set of equations
has three different fixed points~$(\lambda_\sigma^{\ast},\lambda_{\text{V}}^{\ast})$.\footnote{This can be seen by
shifting~$\lambda_{\sigma}\to \lambda_{\sigma}-\lambda_{\text{V}}$ in Eq.~\eqref{eq:dtlvt0}, see Ref.~\cite{Braun:2011pp}.} 
The Gau\ss ian fixed-point at~$(0,0)$ is IR attractive whereas the two non-Gau\ss ian 
fixed-points at $(3 \pi^2, \pi^2)$ and at $(-32 \pi^2 , 16 \pi^2)$ 
have both one IR attractive and one IR repulsive direction, see also Fig.~\ref{fig:flowt0}.

In the following we shall use~$\lambda_\text V = \lambda_\text V ^\parallel = \lambda_\text V ^\perp= 0$ as initial conditions for the couplings associated with
the vector channel interaction, independent of our choice for the temperature and the chemical potential. Thus, these couplings 
are solely induced by quantum fluctuations and do not represent free parameters in our study. In other words, 
the initial value of the scalar-pseudoscalar interaction channel is the only free parameter in our analysis below. Note that this general setup
for the initial conditions of the four-fermion couplings
mimics the situation in many QCD low-energy model studies.
However, since we do not have access to low-energy observables at this order of the derivative expansion,
we shall fix the initial condition of the scalar-pseudoscalar coupling
such that a given value of the critical temperature at vanishing chemical potential is reproduced. 
This determines the scale in our studies of the phase structure below.\footnote{Fixing the critical temperature~$T_{\rm cr}$ to some value at~$\mu=0$
is equivalent to fixing the zero-temperature fermion mass in the IR limit since~$T_{\rm cr}(\mu =0)$ is directly related to the zero-temperature 
fermion mass at~$\mu=0$, at least in a one-channel approximation.} 

For an analysis of the fixed-point structure of our model, the exact value 
of the initial condition of the scalar-pseudoscalar coupling
is not required. Similarly to our discussion of the one-channel approximation, the qualitative features of the ground state of our model 
are already determined by the choice for the initial values of the various couplings relative
to the fixed points.
Provided that the initial value of the scalar-pseudoscalar coupling 
is chosen suitably, i.e. it is 
chosen greater than a critical value~$\lambda_\sigma ^{({\rm cr})}$ depending
on the initial values of the vector-channel couplings,\footnote{Note 
that the function~$\lambda_\sigma ^{({\rm cr})}=\lambda_\sigma ^{({\rm cr})}(\lambda_\text V ^\parallel,\lambda_\text V ^\perp)$
defines a two-dimensional manifold, a separatrix in the space spanned by the couplings of our model. In our one-channel approximation, loosely speaking,
this separatrix is a point which can be identified with {the non-Gau\ss ian fixed-point of} the associated coupling.}
we observe that the four-fermion couplings start to increase rapidly and even diverge at a finite scale~$k_{\rm cr}$, 
indicating the onset of spontaneous symmetry breaking.

In the left panel of Fig.~\ref{fig:flowt0}, an example for an RG trajectory ({pink line}) at zero temperature and chemical potential is shown in the space
spanned by the remaining two couplings~$\lambda_{\sigma}$ and~$\lambda_{\text{V}}$.
In this case the initial condition has been chosen such that the four-fermion couplings diverge at a finite scale~$k_{\rm cr}$. {For~$k\to k_{\rm cr}$, the trajectory 
approaches a separatrix (red line in Fig.~\ref{fig:flowt0}) defining an invariant subspace~\cite{Gehring:2015vja} and indicates a 
dominance of the scalar-pseudoscalar channel, i.e.~$\lambda_\sigma/\lambda_{\text{V}} \approx 3$,
see} also right panel of Fig.~\ref{fig:flowt0} where the RG scale dependence of the two couplings corresponding to this RG trajectory is shown.
This observation appears to be in accordance with the naive expectation that the ground state of our model is governed by spontaneous
chiral symmetry breaking as associated with a dominance of the scalar-pseudoscalar interaction channel. 

The dominance of the scalar-pseudoscalar channel is also observed when finite initial values of
the vector-channel coupling~$\lambda_{\text{V}}$
are chosen, provided that we use a sufficiently large initial value of the scalar-pseudoscalar coupling, see left panel of Fig.~\ref{fig:flowt0}.
However, we would like to emphasize again that this dominance should only be considered as an indicator that the ground state 
in the vacuum limit is governed by chiral symmetry breaking. In particular, our analysis cannot rule out, e.g., 
a possible formation of a vector condensate. For the moment, we shall also leave aside the issue that the {\it Fierz}-complete set of four-fermion interaction channels underlying
this analysis can be transformed into an equivalent {\it Fierz}-complete set of channels with
different transformation properties {regarding the fundamental symmetries} of our model. This further complicates the phenomenological
interpretation, see our discussion of the finite-temperature phase diagram in Sec.~\ref{sec:ps}.

Let us close our discussion of the dynamics of our model in the vacuum limit by commenting on the scaling behavior of the critical
scale~$k_{\text{cr}}$. In the one-channel approximation, we have
found that the scaling of~$k_{\rm cr}$ is of the power-law type with respect to the distance 
of the initial value~$\lambda_{\sigma}^{\text{(UV)}}$ from
the fixed-point value~$\lambda_{\sigma}^{\ast}$, see Eq.~\eqref{eq:plaw}. In our {\it Fierz}-complete setup, this is not necessarily the case. In fact, even
if we set the initial value~$\lambda_{\sigma}^{\text{(UV)}}$ of the scalar-pseudoscalar coupling to zero, the system can still be driven to criticality. This
can be achieved by a sufficiently {large value of the initial condition} of the vector-channel coupling, see left panel of Fig.~\ref{fig:flowt0}.
To be more specific, 
a variation of the vector-channel coupling~$\lambda_{\text{V}}$ in the flow equation~\eqref{eq:dtlst0} of the scalar-pseudoscalar coupling allows to shift
the fixed points of the latter. In particular, a finite value of~$\lambda_{\text{V}}$ turns the Gau\ss ian fixed-point into an interacting fixed point, see Fig.~\ref{fig:es}.
We also deduce from Eq.~\eqref{eq:dtlst0} 
and Fig.~\ref{fig:es} that a critical value~$\lambda_{\text{V}}^{(\text{cr})}$ for the vector-channel coupling 
exists at which the
two fixed points of the~$\lambda_{\sigma}$ coupling merge. For~$\lambda_{\text{V}} > \lambda_{\text{V}}^{(\text{cr})}>0$, the fixed points
of the~$\lambda_{\sigma}$ coupling then annihilate each other and the RG flow is no longer governed by any (finite) real-valued fixed point, resulting
in a diverging~$\lambda_{\sigma}$ coupling. Assuming that the running of the {vector-channel} coupling is sufficiently slow, 
it has been shown~\cite{Braun:2011pp} that the dependence 
of~$k_{\text{cr}}$ on the initial value of the vector coupling obeys a 
{\it Berezinskii-Kosterlitz-Thouless} (BKT) scaling law~\cite{Berezinskii,*Berezinskii2,*Kosterlitz:1973xp},
\be
\!\!\!\!\!\! k_{\text{cr}} \sim \Lambda \theta( \lambda_{\text{V}}^{\text{(UV)}}\!-\! \lambda_{\text{V}}^{(\text{cr})}) \exp\left({-\frac{c_{\text{BKT}}}{\sqrt{\lambda_{\text{V}}^{\text{(UV)}}\!-\! \lambda_{\text{V}}^{(\text{cr})}}} }\right)\,,
\ee
rather than a power law. Here, $c_{\text{BKT}}$ is a positive constant. 
This so-called essential scaling plays a crucial role in gauge theories with many flavors where it is known as {\it Miransky} scaling and the role of our vector coupling
is played by the gauge coupling~\cite{Miransky:1984ef,*Miransky:1988gk,*Miransky:1996pd}. Corrections to this type of scaling behavior arising because of the finite
running of the gauge coupling have found to be of the power-law type~\cite{Braun:2010qs} 
which would translate into corresponding corrections associated with the running of the 
vector coupling in our present study. We emphasize that the dynamics of our present model close to the critical scale is still dominated by the
scalar-pseudoscalar interaction channel in this case, even though the latter has been set to zero initially, 
as can be seen in the flow diagram in the left panel of Fig.~\ref{fig:flowt0}.
\begin{figure}[t]
\begin{center}
  \includegraphics[width=1\linewidth]{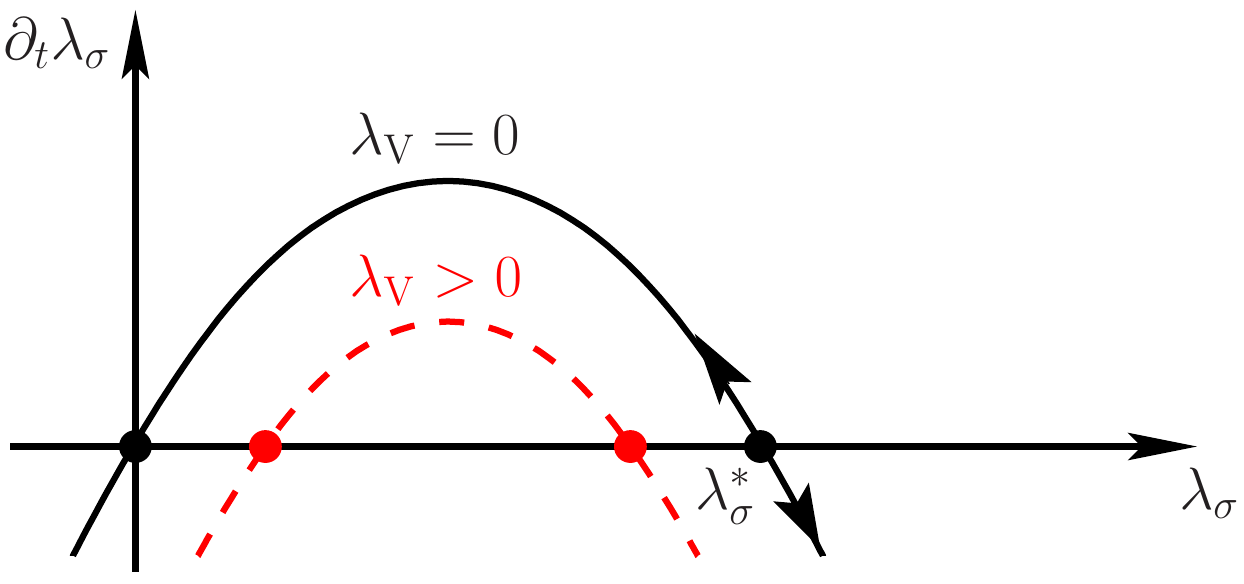}
\end{center}
\caption{Sketch of the $\beta_{\lambda_{\sigma}}$-function of the scalar-pseudoscalar four-fermion coupling for~$\lambda_{\text{V}} = 0$ 
(black line) and~$\lambda_{\text{V}}>0$ (red/dashed
line). The arrows indicate the direction of the RG flow towards the infrared.
}
\label{fig:es}
\end{figure}

A detailed study of the scaling behavior and the associated universality class associated with the quantum phase transitions
potentially occurring in our model in the vacuum limit is beyond the scope of the present work. From now on, we shall rather 
set the vector coupling to zero at the initial scale and let it only be generated dynamically, i.e. we only tune the scalar-pseudoscalar coupling
to fix the scale in our calculations. Still, it is worth mentioning that the mechanism, namely the 
annihilation of fixed points, resulting in
the exponential scaling behavior of~$k_{\rm cr}$ is quite generic. In fact, it also underlies the exponential behavior associated
with the scaling of, e.g., a gap as a function of the chemical potential in case of 
the formation of a {\it Bardeen-Cooper-Schrieffer} (BCS) superfluid in relativistic fermion models. 
We shall discuss the potential occurrence of this type of scaling in more detail in the subsequent section.

\section{Phase structure}\label{sec:ps}

To illustrate the scale-fixing procedure and the computation of
the phase structure at finite temperature and chemical potential, we consider first the approximation with
only a scalar-pseudoscalar interaction channel again. This approximation has also been 
discussed in Refs.~\cite{Braun:2011pp,Aoki:2015hsa}. The results from the 
{\it Fierz}-complete set of flow equations will be discussed subsequently. 

We derive the RG flow
equation for the scalar-pseudoscalar coupling~$\lambda_{\sigma}$ from the full set of flow equations by 
setting~$\lambda_{\text{V}}^{\parallel}=\lambda_{\text{V}}^{\perp}=0$ and also dropping the flow equations associated with these
two couplings, see App.~\ref{app:floweq} for details. 
Moreover, we do not take into account the renormalization of the chemical potential and set~$Z_{\mu}=1$. 
The RG flow equation for $\lambda_\sigma$ then reads
\be
\beta_{\lambda_\sigma} = 2 \lambda_\sigma - 16 v_4 \lambda_\sigma ^2 {\mathcal L}(\tau,\tilde{\mu}_{\tau})\,,
\label{eq:ocappt0m0}
\ee
where~$\tau=T/k$ is the dimensionless temperature,~$\tilde{\mu}_{\tau}=\mu/(2\pi T)$ and
\be
\!\!\!\!\!\! {\mathcal L}(\tau,\tilde{\mu}_{\tau}) &=&
 3 \left(l^\text{(F),(4)}_{\perp +} (\tau, 0,-\I\tilde{\mu}_{\tau}) +  l^\text{(F),(4)}_{\parallel +} (\tau, 0, -\I\tilde{\mu}_{\tau})\right)\nn\\
 && - l^\text{(F),(4)}_{\perp \pm} (\tau, 0, -\I\tilde{\mu}_{\tau}) -  l^\text{(F),(4)}_{\parallel \pm} (\tau, 0,-\I\tilde{\mu}_{\tau})\,.
 \label{eq:lscat}
\ee
The auxiliary function~$\mathcal L$ is simply a sum of so-called threshold functions which essentially represent $1$PI diagrams
describing the decoupling of massive modes and modes in a thermal and/or dense medium. The definition of these
functions can be found in App.~\ref{app:RGtf}. Here, we only note that~${\mathcal L}(0,0)=\frac{1}{2}$. Thus, we recover the
flow equation~\eqref{eq:lsca} in the limit~$T\to 0$ and~$\mu\to 0$.

The flow equation~\eqref{eq:ocappt0m0} for the scalar-pseudoscalar coupling can be solved analytically again. We find
\be
\lambda_{\sigma}(T,\mu,k) = \frac{\lambda_\sigma^\text{(UV)}}{\left(\frac{\Lambda}{k}\right)^{\Theta}
\left(1+ 4\frac{\lambda_\sigma^\text{(UV)}}{\lambda_{\sigma}^{\ast}} {\mathcal I}(T,\mu,k)\right)}\,,
\label{eq:lftsol}
\ee
{where
\be
{\mathcal I}(T,\mu,k) = \frac{1}{\Lambda^2} \int_\Lambda^k {\rm d} k^{\prime} k^{\prime} {\mathcal L}(\tau^{\prime}, \tilde{\mu}_{\tau^{\prime}})
\ee
with} $\tau^{\prime}=T/k^{\prime}$ and~$\lambda_{\sigma}^{\ast}=8\pi^2$ is the non-Gau\ss ian fixed-point value of the scalar-pseudoscalar coupling
at zero temperature and chemical potential, see our discussion in Sec.~\ref{sec:fpsb}.
Using~${\mathcal L}(0,0)=\frac{1}{2}$ to evaluate the solution~\eqref{eq:lftsol} at~$T=\mu=0$, we recover
Eq.~\eqref{eq:sol1c}, as it should be. 

The critical temperature $T_{\rm cr}=T_{\rm cr}(\mu)$ for a given chemical potential~$\mu$ is defined as the temperature at which the 
four-fermion coupling diverges at~$k\to 0$: 
\be
\lim_{k\to 0} \frac{1}{\lambda_{\sigma}(T_{\rm cr},\mu,k)}=0\,.
\ee
With this definition, we obtain the following implicit equation for the critical temperature $T_{\rm cr}$:
\be
0 =  \left(\frac{\lambda_{\sigma}^{\ast}}{\lambda_\sigma^\text{(UV)}}\right) + 4\, {\mathcal I}(T_{\rm cr}, \mu , 0)\,.
\ee
Using Eq.~\eqref{eq:plaw}, we can rewrite this equation in terms of the critical scale~$k_{0}$ at~$T\!=\!\mu\!=\!0$, $k_{0}=k_{\rm cr}(T\!=\!0,\mu\!=\!0)$:
\be
 k_{0} = \Lambda\left(1 + 4\,{\mathcal I}(T_{\rm cr}, \mu , 0)\right)^{\frac{1}{\Theta}}  \,.
\ee
From our discussion of the one-channel approximation in the vacuum limit, it follows immediately that a finite critical
temperature is only found if~$ \lambda_\sigma^\text{(UV)}>\lambda_{\sigma}^{\ast}$.
Apparently, the critical temperature depends on our choice for~$k_{0}$, i.e. on 
the initial condition~$\lambda_{\sigma}^{(\text{UV})}$ of
the scalar-pseudoscalar coupling relative to its fixed-point value. For illustration purposes and
to make a phenomenological connection to QCD, we shall choose a value for the critical temperature at~$\mu=0$ in units of the UV cutoff~$\Lambda$ 
which is close to the chiral critical temperature at~$\mu=0$ found in conventional QCD 
low-energy model studies~\cite{Klevansky:1992qe,Buballa:2003qv,Fukushima:2011jc,Andersen:2014xxa}. 
 To be more specific, we shall fix 
the scale at zero chemical potential 
by tuning the initial condition of the scalar-pseudoscalar coupling such that~$T_0/\Lambda \equiv T_{\rm cr}(\mu=0)/\Lambda=0.15$ 
{and set~$\Lambda=1\,\text{GeV}$}
in the numerical evaluation:
\be
0 = \left(\frac{\lambda_{\sigma}^{\ast}}{\lambda_\sigma^\text{(UV)}}\right) + 4\, {\mathcal I}(T_0 \!=\! 0.15 \Lambda,0,0)\,.
\ee
From here
on, we shall keep the initial condition for the four-fermion coupling fixed to the same value for all 
temperatures and chemical potentials and measure all physical observables in units of~$T_0$.
\begin{figure}[t]
\center
\includegraphics[width=0.5\textwidth]{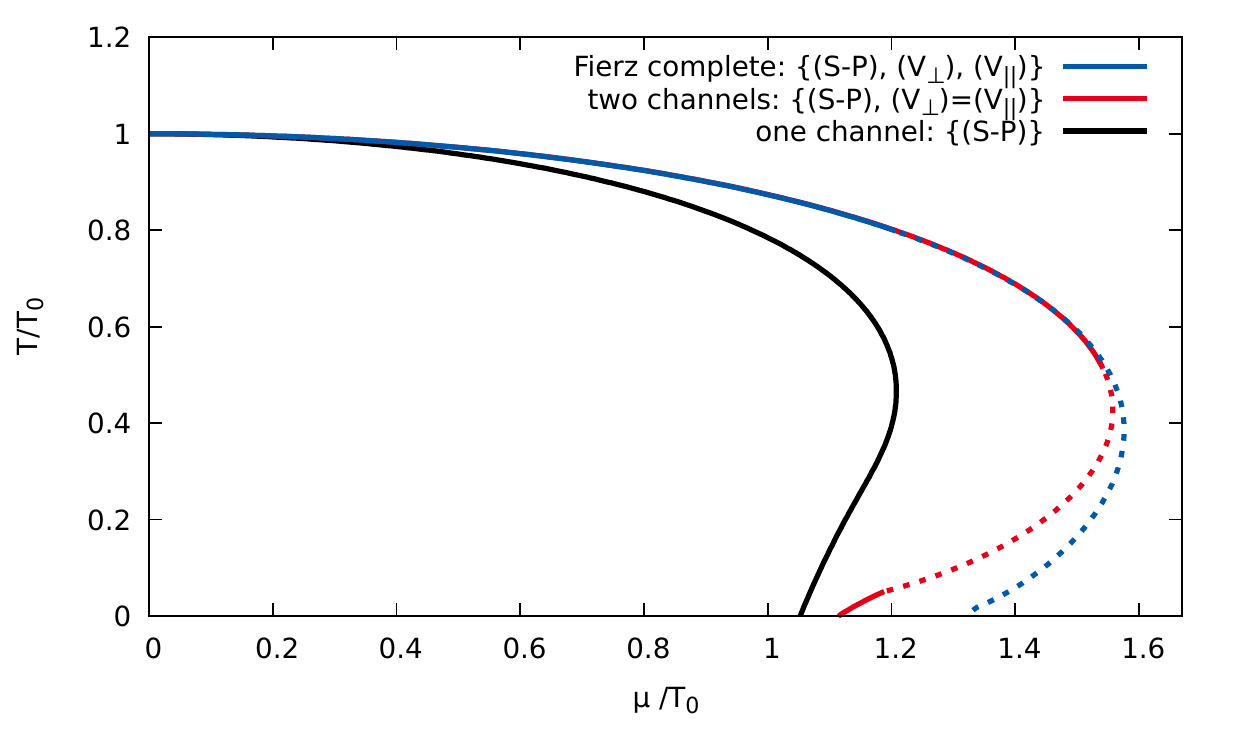}
\caption{Phase boundary associated with the spontaneous breakdown of at least one of the fundamental symmetries of our model
as obtained from a one-channel, two-channel, and {\it Fierz}-complete
study of the ansatz~\eqref{eq:FierzCompleteAnsatz}, see main text for details.} 
\label{fig:ComparisonChannels}
\end{figure}

To ensure comparability of our studies with different numbers of interaction channels, we employ the same 
{scale-fixing procedure} in all cases. As illustrated for the one-channel approximation, we only 
choose a finite value for the initial condition of the scalar-pseudoscalar coupling and
 fix it at zero chemical potential such that the critical temperature
is {given by~$T_0/\Lambda \equiv T_{\rm cr}(\mu=0)/\Lambda=0.15$ in this} limit. The other channels
are only generated dynamically. 
The critical temperature for a given 
chemical potential is still
defined to be the temperature at which the four-fermion couplings diverge at~$k\to 0$. Note that
the structure of the underlying set of flow equations is such that a divergence in one channel
implies a divergence in all interaction channels. However, the various couplings may have 
a different strength relative to each other, see also Fig.~\ref{fig:flowt0} and our discussion in Sec.~\ref{sec:fpsb}.  

In the following we consider the one-channel approximation discussed above, a two-channel
approximation, and the {\it Fierz}-complete system. The RG flow equations 
for the {\it Fierz}-complete set of couplings can be found in App.~\ref{app:floweq}. 
Our two-channel approximation is obtained from this {\it Fierz}-complete set 
by setting~$\lambda_{\text{V}}^{\parallel}=\lambda_{\text{V}}^{\perp}$ and dropping the flow equation of the 
~$\lambda_{\text{V}}^{\parallel}$-coupling. Note that this two-channel approximation is still {\it Fierz}-complete at
zero temperature and chemical potential.

In Fig.~\ref{fig:ComparisonChannels}, we show our results for the~$(T,\mu)$ phase boundary 
associated with the spontaneous breakdown of at least one of the fundamental symmetries of our model. 
We observe right away that
the curvature~$\kappa$ of the finite-temperature phase {boundary,
\be
\kappa = -{T_0}\frac{{\rm d}^2 T_{\rm cr}(\mu)}{{\rm d}\mu^2}\Bigg|_{\mu=0}\,,
\ee
is} significantly smaller in the {\it Fierz}-complete study than in the  
one-channel approximation.\footnote{In order to estimate the curvature, we have
fitted our numerical results for~$T_{\rm cr}(\mu)$ for~$0\leq \mu/T_0 \leq 2/3$ to the 
{ansatz~$T_{\rm cr}(\mu) = T_0 - \frac{1}{2}\kappa \mu^2 + \frac{1}{24}\kappa^{\prime}\mu^4 + {\mathcal O}(\mu^6)$.}}
To be specific,  
the curvature~$\kappa$ in the one-channel approximation is found to be 
{about~44\% greater} than in the {\it Fierz}-complete study. Interestingly, 
the curvature from our two-channel approximation, which is still {\it Fierz}-complete at~$T=\mu=0$,
agrees almost identically with the curvature from the {\it Fierz}-complete {study, see also Tab.~\ref{tab:curvature}.}
\begin{table}[t]
\begin{tabular}{l | c}
\hline\hline
 channels &  curvature $\kappa$ \\
 \hline
$( \text{S} - \text{P})$ & 0.157\\
$( \text{S} - \text{P})$, $(V_{\perp})=(V_{\parallel})$& 0.108\\
{\it Fierz}-complete &  0.109\\
 \hline\hline
\end{tabular}
\caption{\label{tab:curvature}{Curvature~$\kappa$ of the finite-temperature 
phase boundary at~$\mu=0$ as obtained
from a study of a one-channel approximation, a two-channel approximation, and the {\it Fierz}-complete
set of four-fermion channels.  Note that the quoted two-channel approximation is {\it Fierz}-complete 
at~$T=\mu=0$.}} 
\end{table}

From a comparison of the results from the one- and two-channel 
approximation as well as the {\it Fierz}-complete study, we also deduce that the phase boundary
is pushed to larger values of the chemical potential when the number of interaction channels 
is increased. In particular, {we observe that the critical value~$\mu_{\rm cr}$ above which 
{the four-fermion couplings remain finite} 
is pushed to larger values. In fact,~$\mu_{\rm cr}$ 
as obtained from the {\it Fierz}-complete calculation is found
to be 16\% greater than in the two-channel approximation and 20\% greater than in the one-channel approximation.
{Note that~$\mu_{\rm cr}$ is an estimate for the value of the chemical potential
above which no spontaneous symmetry breaking of any kind occurs.}

In addition to these quantitative changes of the phase structure, we observe that the dynamics along
the phase boundary changes on a qualitative level. In the one-channel approximation, the dynamics is completely
dominated by the scalar-pseudoscalar channel by construction. In the two-channel approximation, we then observe
a competition between the scalar-pseudoscalar channel and the vector channel. Indeed, we find that the {vector channel
dominates  
close to} the phase boundary for temperatures~$0.1 \lesssim T/T_0 \lesssim 0.5$, as indicated 
by the red dashed line in Fig.~\ref{fig:ComparisonChannels}. In case of the {\it Fierz}-complete study, we even
observe that the scalar-pseudoscalar channel is only dominant close to the phase boundary for~$T/T_0 \gtrsim 0.8$. 
For~$T/T_0 \lesssim 0.8$, we find a dominance of the $(V_{\parallel})$-channel,
{apart from a small regime~$0.02 \lesssim T/T_0\lesssim 0.09$ in which 
the $(V_{\perp})$-channel {dominates, see also Fig.~\ref{fig:RGFlowT} for an
illustration how the dominance pattern of the channels along the phase boundary changes.}
The dominance of the $(V_{\parallel})$-channel} may not come unexpected 
as it is related to the density, $n \sim \langle \bar{\psi}{\rm i}\gamma_0\psi\rangle$,
which is controlled by the chemical potential.
\begin{figure}[t]
\center
\includegraphics[width=0.5\textwidth]{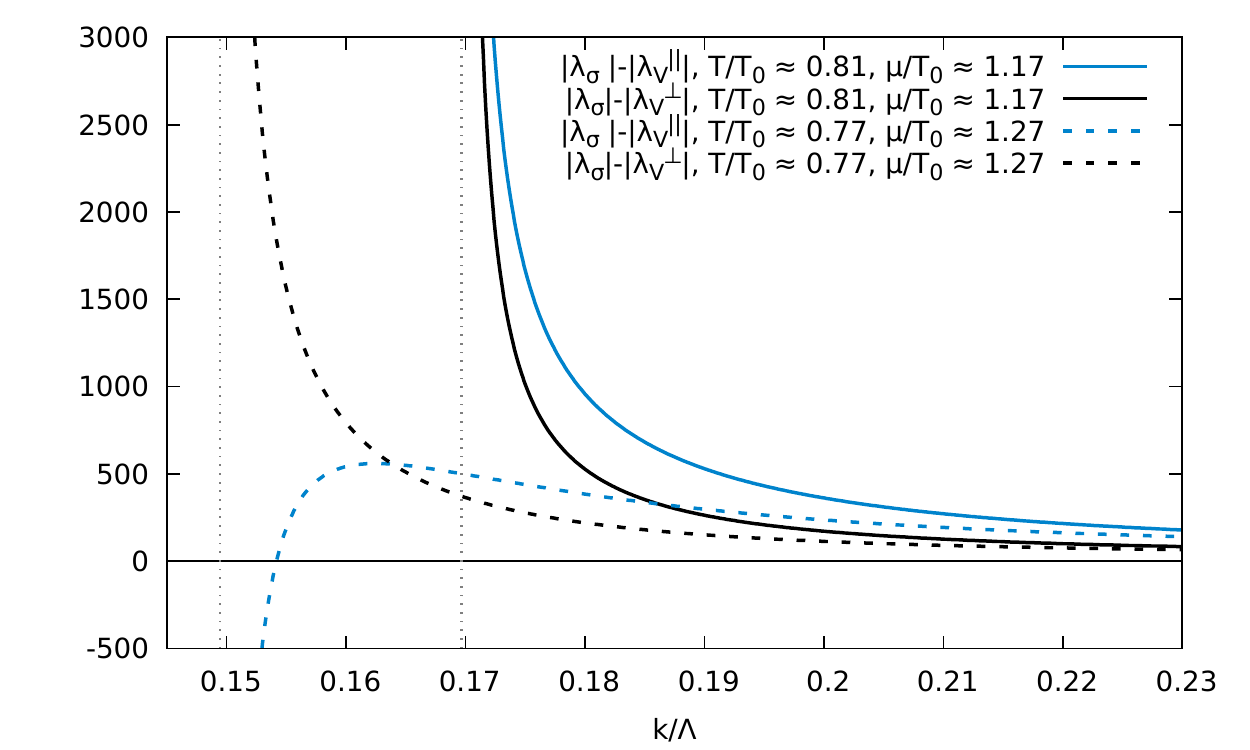}
\caption{RG scale dependence of~$|\lambda_{\sigma}|-|\lambda_{\text{V}}^{\parallel}|$
and~\mbox{$|\lambda_{\sigma}|-|\lambda_{\text{V}}^{\perp}|$} for two sets of values~$(T,\mu)$ corresponding
to two points on the phase boundary associated with the {\it Fierz}-complete study 
shown in Fig.~\ref{fig:ComparisonChannels}. The two points are located closely to the point where the 
dominance pattern of the four-fermion channels changes. From the depicted RG scale dependence 
of the couplings, we indeed deduce that,
at the latter point, the $(V_{\parallel})$-channel starts to dominate over the $\left(\text{S} - \text{P} \right)$-channel while the
$(V_{\perp})$-channel remains to be subdominant. The thin dotted vertical lines indicate
the position of the critical scale~$k_{\rm cr}$ for the chosen values for the temperature and chemical potential.} 
\label{fig:RGFlowT}
\end{figure}

We emphasize again that the dominance of a particular interaction channel only states that the modulus 
of the associated coupling is greater than the ones of the other four-fermion couplings. It does not necessarily imply
that a condensate associated with the most dominant interaction channel is formed. It may therefore only be viewed as an indication
for the formation of such a condensate. Moreover, it may very well be that condensates of different types coexist. 

For example,
note that the dominance of the scalar-pseudoscalar channel close to the phase boundary may be associated
with the formation of a finite chiral condensate, $\varphi \sim \langle \bar{\psi}\psi \rangle$,
which signals the spontaneous breakdown of the chiral $U_{\rm A}(1)$ symmetry of our model. On the other hand, loosely speaking,
a dominance of the $(V_{\parallel})$-channel may be viewed as an indicator for a ``spontaneous breakdown" of {\it Lorentz} invariance in
addition to the inevitable explicit breaking of this invariance introduced by the chemical potential and the temperature.
{A vector-type condensate~$\langle \bar{\psi}\gamma_i\psi\rangle$ associated with a dominance of
the $(V_{\perp})$-channel would furthermore indicate a breakdown of the invariance among the spatial coordinates.
Note that the condensates~$\langle \bar{\psi}{\rm i}\gamma_0\psi\rangle$ and~$\langle \bar{\psi}\gamma_i\psi\rangle$
break neither} the $U_{\rm V}(1)$ symmetry nor the chiral $U_{\rm A}(1)$ symmetry of our model.

The explicit symmetry breaking caused by a finite chemical potential also becomes apparent if we  
introduce an effective density field~$n$ by means of a {\it Hubbard-Stratonovich} transformation. The resulting effective
action then depends on the density~$n$ in form of an explicit field. 
In such a functional, the chemical potential~$\mu$ appears as a term
linear in the density field. The ground state can then be found by solving the quantum equation of motion
in the presence of a finite source being nothing but the {chemical potential, $(\delta \Gamma/\delta n)|_{\mu} = 0$.}
A divergence of the
four-fermion coupling associated with the~$(V_{\parallel})$-channel 
is then related to the coefficient of the~$n^2$-term becoming zero or even negative.
If the appearance of the divergence in the~$(V_{\parallel})$-channel is indeed related to 
a ``spontaneous breakdown" of {\it Lorentz} invariance, then
corresponding pseudo-Goldstone bosons resembling in some aspects a (massive) photon field in temporal gauge
may appear in the spectrum in this regime of the phase diagram.\footnote{Note that our fermionic theory of a single fermion species
may also be viewed as an effective low-energy model for {\it massless} electrons. In QED with {\it massless} electrons (i.e. $U_{\rm A}(1)$-symmetric QED), 
such photon-like pseudo-Goldstone bosons potentially appearing at high densities could mix with the real photons.}    
Symmetry breaking scenarios of this kind have indeed
been discussed in the literature~\cite{Bjorken:1963vg,BialynickiBirula:1963zz,Guralnik:1964zz,Banks:1980rh,oai:arXiv.org:hep-th/0203221}. 
However, their analysis is beyond the scope of the present work. In any case, such a phenomenological interpretation
has to be taken {with some care as we shall see next.}

Our choice for the {\it Fierz}-complete ansatz~\eqref{eq:FierzCompleteAnsatz} is not unique. In order to gain a deeper understanding of
the dynamics of our model and how {\it Fierz}-incomplete approximations may affect the predictive power of model calculations
in general, we consider a second {\it Fierz}-complete parametrization of the four-fermion interaction channels. To this end, 
we introduce explicit difermion channels in our ansatz for the effective {action: 
\be
\Gamma_\text{LO}^\text{(D)} &=&  \int_0 ^\beta d \tau  \int d^3 x \: \bigg \lbrace \psib ( Z^\parallel \I \gamma_0 \del_0 + Z^\perp \I \gamma_i \del_i -  Z_\mu \I \mu \gamma_0 ) \psi \nn \\
& & \hspace{1.4cm} + \frac 1 2 \bar \lambda_{\text{D},\sigma} \left(\text{S} - \text{P} \right) - \frac 1 2 \bar  \lambda _\text{DSP} \left( \text{S} \CC - \text{P} \CC \right)   \nn \\
& &\hspace{2.8cm} - \frac 1 2 \bar  \lambda_\text{D0} \left( \text A _\parallel \CC\right) \bigg \rbrace\,,
\label{eq:dfans}
\ee
where}
\be
\left( \text{S} \CC - \text{P} \CC \right) &\equiv& (\psib \CC \psib ^ T)(\psi^T \CC \psi) - (\psib \gamma_5 \CC \psib^T)(\psi^T \CC \gamma_5 \psi)\,, \nn\\
\left( \text A _\parallel \CC\right) &\equiv&  (\psib \gamma_0 \gamma_5 \CC \psib^T) (\psi^T \CC \gamma_0 \gamma_5 \psi)\,.
\ee
By means of {\it Fierz} transformations (see App.~\ref{App:FierzIdentities}), 
we can rewrite this ansatz
in terms of our original set of interaction channels {introduced in Eq.~\eqref{eq:FierzCompleteAnsatz}:
\be
\Gamma_\text{LO}^\text{(D)} &=&  \int_0 ^\beta d \tau  \int d^3 x \: \bigg \lbrace \psib ( Z^\parallel \I \gamma_0 \del_0 + Z^\perp \I \gamma_i \del_i -  Z_\mu \I \mu \gamma_0 ) \psi \nn \\
& &\hspace{2.2cm} + \frac 1 2 \big( \bar \lambda_{\text{D},\sigma} +  \bar \lambda_\text{DSP} + \frac 1 2 \bar \lambda_\text{D0} \big) \left( \text S - \text P \right) \nn \\
& &\hspace{2.2cm}- \frac 1 2 \big( - \bar \lambda_\text{DSP} -\frac 3 2 \bar \lambda _\text{D0} \big) \left( V_\parallel \right) \nn \\
& &\hspace{2.2cm}- \frac 1 2 \big( - \bar \lambda _\text{DSP} + \frac 1 2 \bar \lambda_\text{D0} \big) \left( V_\perp \right) \bigg \rbrace,
\ee
This} allows us to identify the following relations between the various {couplings:
\be
\bar \lambda_\sigma &=&  \bar \lambda_{\text{D},\sigma} +  \bar \lambda_\text{DSP} + \frac 1 2 \bar \lambda_\text{D0}\,,  
\label{eq:RelationCouplings1}\\
%
%
\bar \lambda_\text V ^\parallel &=&  - \bar \lambda_\text{DSP} -\frac 3 2 \bar \lambda _\text{D0}\,,
\label{eq:RelationCouplings2}\\
%
%
\bar \lambda_\text V ^\perp &=& - \bar \lambda _\text{DSP} + \frac 1 2 \bar \lambda_\text{D0}\,.
 \label{eq:RelationCouplings3}
\ee
By} inverting these relations we eventually obtain the $\beta$ functions of the couplings in our 
``difermion parametrization" of the effective {action:
\be
\del_t  \lambda_{\text{D},\sigma} &=& \beta _{\lambda_\sigma} +\frac 1 2  \beta_{\lambda_\text V ^\parallel} +\frac 1 2  \beta_{\lambda_\text V ^\perp}, \\
\del_t  \lambda_\text{DSP} &=& - \frac 1 4 \beta_{\lambda_\text V ^\parallel} - \frac 3 4 \beta _{\lambda_\text V ^\perp}, \\
\del_t  \lambda_\text{D0} &=& - \frac 1 2 \beta_{\lambda_\text V ^\parallel} + \frac 1 2 \beta_{\lambda _\text V^\perp}.
\label{eq:dffloweq}
\ee
The} $\beta$ functions on the right-hand side depend on the couplings $\{ \bar \lambda_\sigma, \bar \lambda_\text V ^ \parallel, \bar \lambda_\text V ^\perp \}$
and can be expressed in terms of the 
couplings~$\{  \bar \lambda_{{\rm D},\sigma}, \bar \lambda _\text{DSP}, \bar \lambda_\text{D0}\}$ 
using Eqs.~\eqref{eq:RelationCouplings1}-\eqref{eq:RelationCouplings3}.
Note that, at~$T=\mu=0$, the flow of the~$\lambda_\text{DSP}$ coupling is {up to a global minus sign 
identical to the flow of the vector coupling~${\lambda}_{\text{V}}$ in the effective action~\eqref{eq:FierzCompleteAnsatz}.}

The $\left(\text{S} - \text{P} \right)$-channel in our ansatz~\eqref{eq:dfans} is again the conventional scalar-pseudoscalar channel. 
A dominance of this channel indicates the onset of spontaneous chiral~$U_{\rm A}(1)$ symmetry breaking in our model. A dominance
of the difermion channel~$( \text{S} \CC - \text{P} \CC)$ is associated with the spontaneous breakdown of 
both the chiral~$U_{\rm A}(1)$ symmetry and the~$U_{\rm V}(1)$ symmetry of our model. Thus, a 
dominance of the~$( \text{S} \CC - \text{P} \CC )$-channel also suggests chiral symmetry breaking as measured 
by the conventional~$\left(\text{S} - \text{P} \right)$-channel and, loosely speaking, the information encoded in both
channels is therefore not disjunct. In contrast to our 
previous ansatz~\eqref{eq:FierzCompleteAnsatz}, however, the parametrization of the four-fermion couplings
in the ansatz~\eqref{eq:dfans} 
allows to probe more directly a possible spontaneous breakdown
of the~$U_{\rm V}(1)$ symmetry. 
Phenomenologically, 
the latter may naively be associated with  
the formation of a BCS-type superfluid ground state. In particular, a dominance
of this channel may indicate the formation of a 
finite difermion condensate~$\langle\psi^{T}\CC\gamma_5\psi\rangle$ in the 
scalar~$J^{P}=0^{+}$ channel.\footnote{Note that it is not possible in our present model to construct 
a {\it Poincar\'{e}}-invariant~$J^{P}=0^{+}$ condensation 
channel (from a corresponding four-fermion interaction channel) which only breaks~$U_{\rm V}(1)$ symmetry but leaves the 
chiral~$U_{\rm A}(1)$ symmetry intact. In QCD, the formation of the associated 
diquark condensate can be realized at the price of a broken $SU(3)$ color symmetry, even if the chiral symmetry remains unbroken. 
In QED, on the other hand, the required
breaking of the~$U_{\rm A}(1)$ symmetry is realized by a finite explicit electron mass.} 
We emphasize that these considerations do not imply that the ansatz~\eqref{eq:dfans} is more general by any means. In fact, as we have shown,
both ans\"atze are equivalent as they are related by {\it Fierz} transformations. Therefore, {these considerations 
only make obvious that the potential
formation of a~$U_{\rm V}(1)$-breaking ground state may just not be directly visible in a study with 
the ansatz~\eqref{eq:FierzCompleteAnsatz} but may nevertheless be realized by 
a specific simultaneous formation of two condensates, 
namely a~$\langle \bar{\psi}\gamma_0\psi\rangle$ condensate 
and a~$\langle \bar{\psi}\gamma_i\psi\rangle$ condensate, according to 
Eqs.~\eqref{eq:RelationCouplings1}-\eqref{eq:dffloweq}.\footnote{Within a truncated 
bosonized formulation (e.g. mean-field approximation), the specific choice for the parametrization of the
four-fermion interaction channels is of great importance as it determines the choice for the associated 
bosonic fields (e.g. mean fields). The latter effectively determine a specific parametrization
of the momentum dependence of the four-fermion channels. Therefore, the parametrization of the action in terms of 
four-fermion channels is of relevance from a phenomenological
point of view. To be specific, even if two actions are equivalent on the level of {\it Fierz} transformations, the results from
the mean-field studies associated with the two actions will in general be different.}
We} add that a dominance of the~$( \text A _\parallel \CC )$-channel may indicate the formation of a 
condensate~$\langle\psi^T \CC \gamma_0 \gamma_5 \psi\rangle$ with positive parity
which breaks the~$U_{\rm V}(1)$ symmetry of our model but leaves the chiral~$U_{\rm A}(1)$ symmetry intact. However, this channel
also breaks explicitly {\it Poincar\'{e}} invariance.
\begin{figure}[t]
\centering
\includegraphics[width=0.5\textwidth]{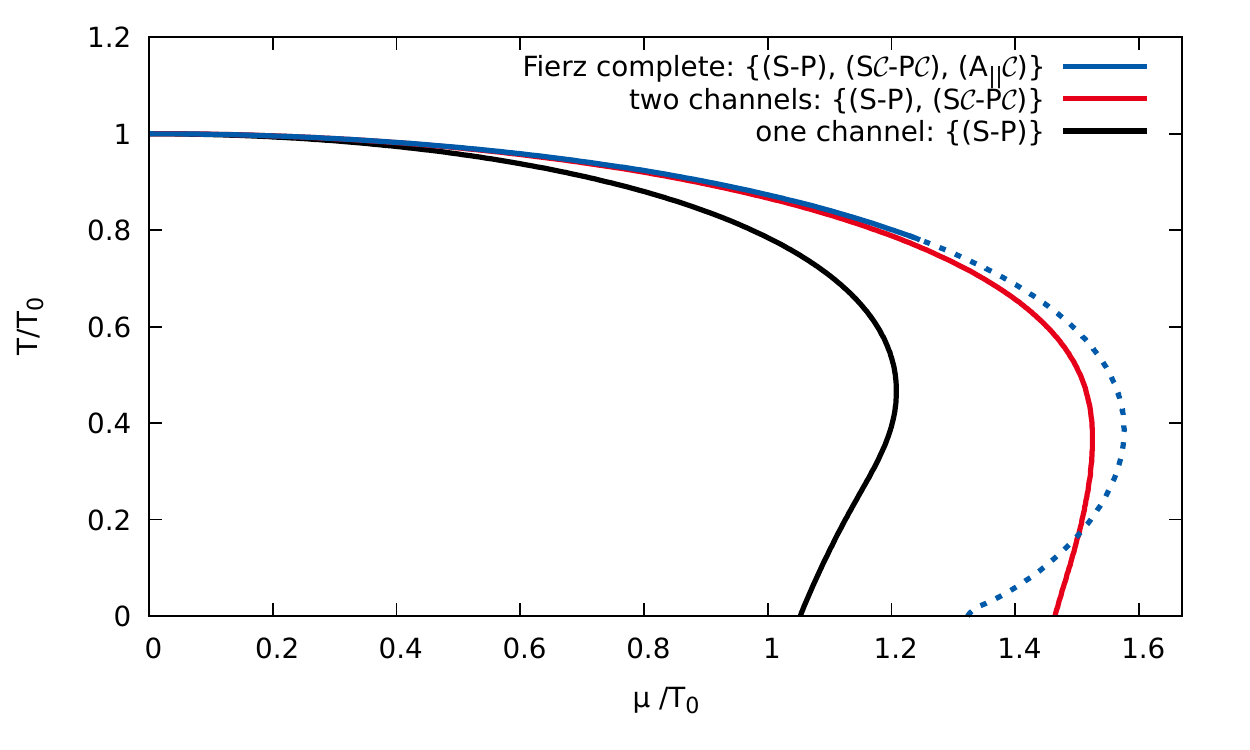}
\caption{Phase boundary associated with the spontaneous breakdown of at least one of the 
fundamental symmetries of our model as obtained from a one-channel, two-channel, and Fierz-complete study of
the ansatz~\eqref{eq:dfans}, see main text for details.}
\label{fig:dfpd}
\end{figure}

From our comparison of the ans\"atze~\eqref{eq:FierzCompleteAnsatz} and~\eqref{eq:dfans}, 
we immediately conclude that a phenomenological interpretation of the symmetry breaking patterns of our model 
requires great care. This is even more the case when a {\it Fierz}-incomplete set of four-fermion interactions is considered 
which has been extracted from a specific {\it Fierz}-complete parametrization of the interaction channels.
\begin{figure*}[t]
\begin{center}
  \includegraphics[width=0.47\linewidth]{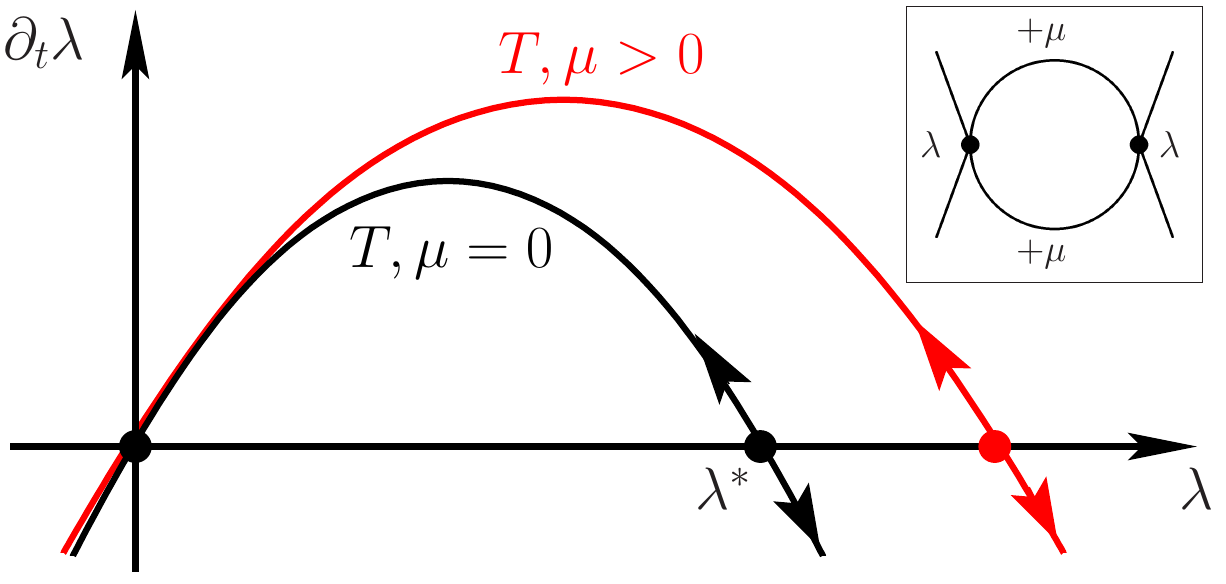}
   \hfill
\includegraphics[width=0.47\linewidth]{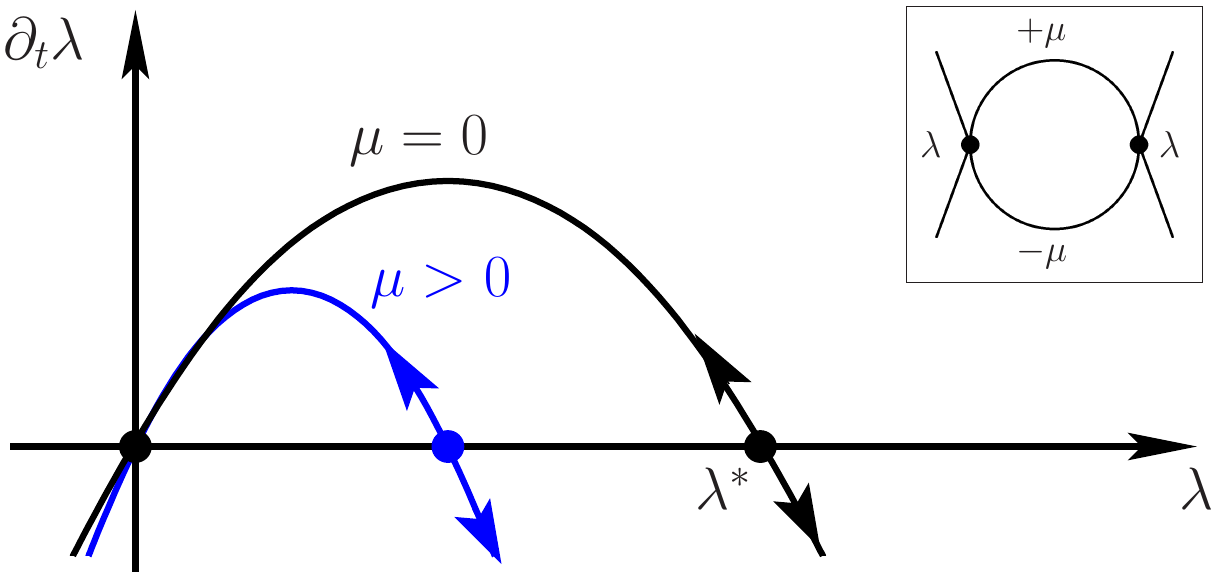}
\end{center}
\caption{Left panel: Sketch of the $\beta$-function of a four-fermion coupling which is only driven by a diagram of the type 
as shown in the inset. For increasing~$T/k$ and~$\mu/k$, the non-Gau\ss ian fixed-point is shifted to larger values.
Right panel:  Sketch of the $\beta$-function of a four-fermion coupling at~$T=0$ which is only driven by a diagram of the type
as shown in the inset. For increasing~$\mu/k$, the non-Gau\ss ian fixed-point is shifted to smaller values and eventually merges
with the Gau\ss ian fixed-point.
}
\label{fig:pmpp}
\end{figure*}

In Fig.~\ref{fig:dfpd}, we show our results for the~$(T,\mu)$ phase boundary 
associated with the spontaneous breakdown of at least one of the fundamental symmetries of our model which are 
now encoded in the four-fermion interaction channels
as parametrized in our ansatz~\eqref{eq:dfans} for the effective action. The one-channel approximation is the same
as in the {case of our ansatz~\eqref{eq:FierzCompleteAnsatz} for} the effective action and the results for the phase boundary (solid black line) 
are only shown to guide the eye. Moreover, the location 
of the phase boundary from the {\it Fierz}-complete study of the effective action~\eqref{eq:dfans}
agrees identically with the {\it Fierz}-complete study of the effective action~\eqref{eq:dfans}, as it should be.
{In the present case, we observe again a dominance of the $(\text{S} - \text{P})$-channel 
close to the phase boundary 
for {temperatures~$1 \geq T/T_0 \gtrsim 0.8$ (solid blue line in Fig.~\ref{fig:dfpd})}.
In the light of our results from the parametrization~\eqref{eq:FierzCompleteAnsatz} of 
the effective action, where the $(\text{S}-\text{P})$-channel has also been found to be dominant close to the phase boundary
for~$1 \geq T/T_0 \gtrsim 0.8$, 
we may now cautiously conclude from the combination of the results from the two ans\"atze
that at least the phase boundary in the temperature regime~$1 \geq T/T_0 \gtrsim 0.8$ is}
associated with spontaneous chiral symmetry breaking as the latter is indicated by 
a dominance of either the $\left(\text{S} - \text{P} \right)$-channel
or the $( \text{S} \CC - \text{P} \CC)$-channel.

{In line with our study based on the parametrization~\eqref{eq:FierzCompleteAnsatz} of 
the effective action, we now also observe a dominance of a channel associated with broken {\it Poincar\'{e}}
invariance at~$0\leq T/T_0\lesssim 0.8$ in} the {\it Fierz}-complete study (dashed blue line in Fig.~\ref{fig:dfpd}),
namely a dominance of the $( \text A _\parallel \CC )$-channel. 
In case of the two-channel approximation, which has been obtained by setting~$\bar{\lambda}_{\text{D0}}=0$ and dropping the 
corresponding flow equation, {we only observe a dominance of the 
$(\text{S}-\text{P})$-channel (solid red line) close to the phase boundary for all temperatures~$1\geq T/T_0 \geq 0$.}

From a comparison of the results from the one- and two-channel 
approximation, we also deduce that the phase boundary
is again pushed to larger values of the chemical potential. However, we now observe that the phase boundary is 
pushed back to smaller values of the chemical potential again {at low temperature} when we go from the two-channel approximation
to the {\it Fierz}-complete ansatz. This underscores again that a phenomenological interpretation of the phase structure and
symmetry breaking patterns in {\it Fierz}-incomplete studies have to be taken with some care.

Whereas the phenomenological interpretation 
of the dominance of the various interaction channels in different parametrizations
of the effective action may be difficult, a qualitative insight into the symmetry breaking mechanisms
can be obtained from an analysis of the fixed-point structure of the four-fermion couplings. To this end, we may consider
the temperature and the chemical potential as external couplings, governed by a trivial dimensional RG running~$\partial_t (T/k) = -(T/k)$
and~$\partial_t (\mu/k) = -(\mu/k)$.

Two types of diagrams essentially contribute to the RG flow of the four-fermion couplings at finite chemical potential, see insets
of Fig.~\ref{fig:pmpp} for diagrammatic representations and App.~\ref{app:floweq} for explicit expressions of the flow equations. 
In a partially bosonized formulation
of our model, the interaction between the fermions 
is mediated by the exchange of bosons with fermion number~$F=0$ and zero chemical potential 
(corresponding to states with zero baryon number in QCD, such as pions) 
in the diagram in the inset of the left panel of Fig.~\ref{fig:pmpp}. On the other hand,
the fermion interaction is mediated
by a bosonic difermion state with fermion number~$|F|=2$ and an effective chemical potential~$\mu_{\rm D}=F\mu$
in the diagram in the inset of the right panel of Fig.~\ref{fig:pmpp}.

Let us now assume that the RG flow of a given four-fermion coupling~$\lambda$ is only governed by diagrams of
the type shown in the inset of the left panel of Fig.~\ref{fig:pmpp}. The RG flow equation is then given by
\be
\partial _t \lambda = 2\lambda - c_{+}l^{\text{(F)}}_{+}\lambda^2\,,
\ee
where, without loss of generality, we assume~$c_{+}$ is a positive numerical constant. This flow
equation has a Gau\ss ian fixed-point and a non-Gau\ss ian fixed-point~$\lambda^{\ast}$. Strictly 
speaking, the latter becomes a {pseudo fixed-point in the presence} of an external parameter, such as a
finite temperature and/or finite chemical potential.
The so-called threshold function~$l^{\text{(F)}}_{+}$ depends on the dimensionless
temperature~$T/k$ as well as the dimensionless chemical potential~$\mu/k$ and 
essentially represents the loop diagram in 
the inset of the left panel of Fig.~\ref{fig:pmpp}. For an explicit representation of such a threshold 
function, we refer the reader to App.~\ref{app:RGtf}. Note that all threshold functions in this work come in two different
variations, e.g.~$l^{\text{(F)}}_{\parallel,+}$ and~$l^{\text{(F)}}_{\perp,+}$, which can be traced back to the tensor structure
becoming more involved due to the explicit breaking of {\it Poincar\'{e}} invariance. Although we have taken this into account
in our numerical studies, we leave this subtlety aside in our more qualitative discussion at this point. 

For increasing dimensionless temperature~$T/k$ 
at fixed dimensionless chemical potential~$\mu/k$, we have  
\be
l^{\text{(F)}}_{+}\to 0\quad\text{for}\quad {\frac{T}{k}} \gg 1\,,
\ee
due to the thermal screening of the fermionic
modes. Moreover, we also have~$l^{\text{(F)}}_{+} \to 0$ for sufficiently large values 
of~$\mu/k$ for a given fixed dimensionless temperature~$T/k$. This implies that the fermions become effectively weakly interacting
in the dense limit. Overall, we have~$\lambda^{\ast}\to\infty$ for the 
non-Gau\ss ian fixed-point 
for~$T/k\to \infty$ and/or~$\mu/k\to\infty$, see left panel of Fig.~\ref{fig:pmpp}. 
Let us now assume that we have fixed the initial condition~$\lambda^{(\rm UV)}$ of the
four-fermion coupling such that~$\lambda^{(\rm UV)} > \lambda^{\ast}$ at~$T=0$ and~$\mu=0$ and
keep it fixed to the same value for all values of~$T$ and~$\mu$.
As discussed in detail above, the four-fermion
coupling at~$T=0$ and~$\mu=0$ then increases rapidly towards the IR, indicating the onset of spontaneous symmetry breaking. 
However, since the value of the non-Gau\ss ian fixed-point increases with increasing~$T/k$ and/or 
increasing~$\mu/k$, the rapid increase of the four-fermion coupling towards the IR is effectively slowed down 
and may even change its direction in the space defined by the coupling~$\lambda$, the dimensionless temperature~$T/k$
and the dimensionless chemical potential~$\mu/k$.
This behavior of the (pseudo) non-Gau\ss ian fixed-point suggests that, for a fixed initial value~$\lambda^{(\rm UV)}>\lambda^{\ast}$,
a critical temperature~$T_{\rm cr}$ as well as a critical chemical potential~$\mu_{\rm cr}$
exist above which the four-fermion coupling 
does not diverge but approaches zero in the IR and therefore the symmetry associated with the coupling~$\lambda$
is restored. At least at high temperature, 
such a behavior is indeed expected since the fermions become effectively ``stiff" 
degrees of freedom due to their thermal Matsubara mass~$\sim T$. This is the type of symmetry restoration mechanism 
which dominantly determines the phase structure of our model at finite temperature and chemical potential, as indicated
in Figs.~\ref{fig:ComparisonChannels} and~\ref{fig:dfpd}
by the finite extent of the regime associated with spontaneous symmetry breaking in both~$T$- and~$\mu$-direction.
We may even cautiously deduce from this observation that the dynamics {close to 
and below the phase boundary at low temperature}
is governed by the formation of a condensate with fermion number~$F=0$ as the general structure of the phase diagram
appears to be dominated by diagrams of the type shown in the inset of the left panel of Fig.~\ref{fig:pmpp}.

A dominance of the RG flow by diagrams of the type 
shown in the inset of the right panel of Fig.~\ref{fig:pmpp} 
would suggest the formation of a condensate with fermion number~$|F|=2$, i.e. a difermion-type condensate.
In this case, we would indeed expect a different phase structure, at least at (very) low temperature and large chemical potential.
To illustrate this, let us now assume that the RG flow of a given four-fermion coupling~$\lambda$ is only governed by diagrams of
the form shown in the inset of the right panel of Fig.~\ref{fig:pmpp}:
\be
\partial _t \lambda = 2\lambda - c_{\pm}l^{\text{(F)}}_{\pm}\lambda^2\,,\label{eq:cisfe}
\ee
where, again without loss of generality, we assume~$c_{\pm}$ is a positive numerical constant. 
This flow
equation has a Gau\ss ian fixed-point and a non-Gau\ss ian fixed-point~$\lambda^{\ast}$. 
The so-called threshold function~$l^{\text{(F)}}_{\pm}$ depends on the (dimensionless)
temperature~$T/k$ and the dimensionless chemical potential~$\mu/k$ and represents the 
associated loop integral. 
Explicit representations of this type of threshold 
function can be found in App.~\ref{app:RGtf}. 
For increasing~$T/k$ at fixed~$\mu/k$, we find again 
\be
l^{\text{(F)}}_{\pm}\to 0\quad\text{for}\quad {\frac{T}{k}} \gg 1\,,
\ee
due to the thermal screening of the fermionic
modes. However, we have  
\be
l^{\text{(F)}}_{\pm} \sim \left(\frac{\mu}{k}\right)^2\quad\text{for}\quad \frac{\mu}{k} \gg 1\,
\ee
at~$T=0$. For finite fixed~$T/k$, we then observe that~$l^{\text{(F)}}_{\pm}$ 
increases as a function of~$\mu/k$ until it reaches a maximum and
then tends to zero for~$\mu/k\to\infty$. The position of the maximum
is shifted to smaller values of~$\mu/k$ for increasing~$T/k$. 

Let us now focus
on the strict zero-temperature limit. In this case, the value of the non-Gau\ss ian fixed-point is decreased for
increasing~$\mu/k$ and eventually merges with the Gau\ss ian fixed-point. This implies immediately that
the four-fermion coupling always increases rapidly towards the IR for~$\mu>0$, indicating the onset of
spontaneous symmetry breaking, provided that the initial condition~$\lambda^{(\rm UV)}$
has been chosen positive,~$\lambda^{(\rm UV)}>0$.\footnote{For~$c_{\pm}<0$, $\lambda^{(\rm UV)}$ has 
to be chosen negative in order to trigger spontaneous symmetry breaking in the long-range limit.}
Thus, the actual choice for~$\lambda^{(\rm UV)}$ relative to the 
value of the non-Gau\ss ian fixed-point plays a less prominent
role in this case, at least on a qualitative level. In other words, an infinitesimally small positive coupling triggers 
the formation of a condensate with fermion number~$|F|=2$. This is nothing but the {\it Cooper} instability
in the presence of an arbitrarily weak attraction~\cite{Cooper:1956zz} which destabilizes the {\it Fermi} sphere and
results in the formation of a {\it Cooper} pair condensate~\cite{Bardeen:1957kj,Bardeen:1957mv}, inducing a gap in the excitation spectrum.
For~$\lambda^{(\rm UV)}=0$, the four-fermion coupling remains at the Gau\ss ian fixed. For~$\lambda^{(\rm UV)}<0$, 
the theory approaches the Gau\ss ian fixed-point in the IR limit. Thus, there is no spontaneous symmetry breaking 
for~$\lambda^{(\rm UV)}\leq 0$.  

The fact that the two fixed points merge for~$\mu/k\to \infty$ at~$T=0$ leaves its imprint in the $\mu$-dependence
of the critical scale~$k_{\rm cr}$ at which the four-fermion coupling diverges.
In fact, from the flow equation~\eqref{eq:cisfe}, we recover the typical BCS-type exponential scaling behavior of the critical scale:
\be
k_{\rm cr} = \Lambda_0\theta(\bar{\lambda}_0)\exp\left(-\frac{c_{\text{BCS}}}{\mu^2\bar{\lambda}_{0}}\right)\,.
\label{eq:BCSsb}
\ee
Here, we have assumed that the RG flow equation~\eqref{eq:cisfe} has been initialized in the IR regime at~$k =\Lambda_0<\Lambda$ 
with an initial value~$\bar{\lambda}_0>0$,
such that~$l^{\text{(F)}}_{\pm}$ can be approximated by 
$l^{\text{(F)}}_{\pm} = c_{\infty} (\mu/k)^2$ with $c_{\infty} >0$. 
Moreover, we have introduced the numerical
constant~$c_{\text{BCS}}= 1/(c_{\infty} c_{\pm}) > 0$. 
The value~$\bar{\lambda}_0$ of
the four-fermion coupling can be directly related to the UV coupling~$\lambda^{(\rm UV)}$.
Recall that the dependence 
of~$k_{\rm cr}$ on the chemical potential is then handed down to 
physical observables in the infrared limit, 
leading to the typical exponential scaling behavior~\cite{Bailin:1983bm}.

The observed exponential-type scaling behavior of the scale~$k_{\rm cr}$
appears to be generic in cases where two fixed points merge, see, e.g., our discussion of essential scaling (BKT-type scaling) in
Sec.~\ref{sec:fpsb} which plays a crucial role in gauge theories with many 
flavors~\cite{Miransky:1984ef,Miransky:1988gk,Miransky:1996pd,Braun:2010qs}. 

At finite temperature
and chemical potential, the shift of the
non-Gau\ss ian fixed-point towards the Gau\ss ian fixed-point is slowed down and eventually inverted such that the value of the 
non-Gau\ss ian fixed-point eventually increases with increasing~$T/k$. 
This suggests again that a critical temperature 
exists above which the symmetry associated with 
the coupling~$\lambda$ is restored.

If the ground state of our model at large chemical potential was governed by the {\it Cooper} instability as associated with the 
exponential scaling behavior~\eqref{eq:BCSsb} 
of the scale~$k_{\rm cr}$, then the phase boundary would extend to arbitrarily large values of
the chemical potential, at least in the strict zero-temperature limit. 
However, this is not observed in the numerical solution
of the full set of RG flow equations, see Figs.~\ref{fig:ComparisonChannels} and~\ref{fig:dfpd}. 
Of course, this does not imply
that {difermion-type phases are not favored at all in this model (e.g. a phase with 
a chirally invariant $U_{\rm V}(1)$-breaking $\langle\psi^T \CC \gamma_0 \gamma_5 \psi\rangle$-condensate)} since the phase structure also depends on our choice for the
initial conditions of the four-fermion couplings. The formation of such phases may therefore be realized by a suitable tuning
of the initial conditions. Still, the vacuum phase structure of our model suggests that the general features of the
phase diagrams presented in Figs.~\ref{fig:ComparisonChannels} and~\ref{fig:dfpd} persist over a significant range of initial
values for the couplings~$\lambda_{\sigma}$ and~$\lambda_{\rm V}$, see Fig.~\ref{fig:flowt0}.

\section{Conclusions}\label{sec:conc}
In this work we have analyzed the phase structure of a one-flavor NJL model at finite temperature 
and chemical potential. With the aid of RG flow equations, we aimed at an understanding on how
{\it Fierz}-incomplete approximations affect the predictive power of general NJL-type models, which are also 
frequently employed to study the phase structure of QCD.
To this end, we have considered the RG flow of four-fermion couplings at leading order 
of the derivative expansion. This approximation
already includes corrections beyond the mean-field approximation which is inevitable to preserve the invariance
of the results under {\it Fierz} transformations. 

We have found {that {\it Fierz}-incompleteness affects strongly} key quantities, such as the curvature of the phase boundary
at small chemical potential. Indeed, the curvature obtained in a calculation including only the conventional scalar-pseudoscalar 
channel has been found to {be about~$44\%$ greater} than in the {\it Fierz}-complete study. With respect to the 
critical value~$\mu_{\rm cr}$ 
of the chemical potential above which no spontaneous symmetry breaking occurs, we have found that~$\mu_{\rm cr}$ in 
the {\it Fierz}-complete study is about~$20\%$ greater than in the conventional one-channel approximation.
Moreover, we have observed
that the position of the phase boundary depends strongly on the number of four-fermion channels 
included in {\it Fierz}-incomplete studies.
In general, {\it Fierz}-incomplete calculations may either
overestimate or underestimate the size of the regime governed by spontaneous symmetry breaking
in the~$(T,\mu)$ plane. 
The actual approach to the result from the {\it Fierz}-complete
study depends strongly on the type of the channels included in such studies.
In fact,
our analysis suggests that
the use of {\it Fierz}-incomplete approximations may even lead to the prediction of spurious phases, in particular at large chemical
potential.

With respect to a determination of the properties of the actual {ground 
state in the phase governed by spontaneous symmetry breaking}, 
our present study based on the analysis of RG flow equations at leading order of the derivative expansion is limited. In order
to gain at least some insight into this question, we have analyzed which four-fermion channel dominates the dynamics of the system 
close to the phase boundary. A dominance of a given channel may then indicate the formation of 
a corresponding condensate. As we have discussed, 
however, this criterion has to be taken with some care, in particular when only one specific parametrization of the four-fermion
channels is considered. This also holds for {\it Fierz}-complete studies. In this work, we have used two different 
{\it Fierz}-complete parametrizations
and found that, over a wide range of the chemical potential, the dynamics close to the phase boundary is dominated 
by {the conventional scalar-pseudoscalar channel associated with 
chiral symmetry breaking. At large chemical potential, the dynamics
close to the phase boundary then appears in both cases to be dominated by channels which break explicitly {\it Poincar\'{e}} invariance.}

As a second criterion for the determination of at least some properties of the ground state of the regime governed by
spontaneous symmetry breaking, we have analyzed on general grounds the scaling behavior of the loop diagrams contributing
to the RG flow of the four-fermion couplings. The scaling of these diagrams as a function of the dimensionless temperature and
chemical potential determines the fixed-point structure of the theory at finite temperature and chemical potential. Our
fixed-point analysis suggests {that the dynamics close} to and below the phase boundary
is governed by the formation of a condensate with fermion number~$F=0$. In 
contrast to QCD (see, e.g., Refs.~\cite{Bailin:1983bm,Buballa:2003qv,Alford:2007xm,Anglani:2013gfu} for reviews), 
the formation of difermion-type condensates with fermion number~$|F|=2$ does not appear to {be 
favored, at} least at large chemical potential 
for the initial conditions of the RG flow equations employed in our present study.

Of course, at the present order of the derivative expansion, 
the employed criteria can only serve as indicators for the actual properties of the ground state in regimes 
governed by spontaneous symmetry breaking. In order to determine
the ground-state properties of a system unambiguously, a calculation of the full order-parameter potential is eventually required. Still,
we expect that our present analysis may already turn out to be useful to guide future NJL-type model studies.

{\it Acknowledgments.--~} The authors {would like to thank H.~Gies, J.~M.~Pawlowski, and D.~Roscher}
for useful discussions and comments on the manuscript. 
In addition, the authors are grateful to M.~Q.~Huber and F.~Karbstein for correspondence and, 
as members of the {\it fQCD collaboration}~\cite{fQCD}, the authors also would like to 
thank the other members of this collaboration for discussions.
J.B. acknowledges support by HIC for FAIR within the LOEWE program of the State of Hesse. 
This work is supported by the DFG through grant SFB 1245.


\appendix

\section{Fierz Identities}\label{App:FierzIdentities}
We have used the following {\it Fierz} identities to derive Eq.~\eqref{eq:FierzCompleteAnsatz} {from Eq.~\eqref{eq:gammagen}: 
\be
(\text{A}_{\parallel})  &=& \frac{1}{2} (\text{S}-\text{P}) - \frac{1}{2}\left(\text{V}_{\parallel}\right) + \frac{1}{2} \left(\text{V}_{\perp}\right)\,,   \\
(\text{A}_{\perp}) &=&  \frac{3}{2} (\text{S}-\text{P})   + \frac{3}{2} \left(\text{V}_{\parallel}\right) + \frac{1}{2}\left(\text{V}_{\perp}\right)\,,   \\
\left(\text{T}_{\parallel}\right)  &=& 3\left(\text{V}_{\parallel}\right) -  \left(\text{V}_{\perp}\right)\,.
\ee
The} {\it Fierz} transformations from the fermion-antifermion channels to the difermion-type 
channels are given by
\be
\left( \text S \CC - \text P \CC \right) &=& - \left( \text S - \text P \right) - \left( \text V_\parallel \right) - \left( \text V _\perp \right)\,, \\
\left( \text A _\parallel \CC \right) &=& -\frac{1}{2} \left( \text S - \text P \right) -\frac{3}{2}  \left( \text V_\parallel \right) + \frac{1}{2} \left( \text V _\perp \right)\,, \\
\left( \text A _\perp \CC \right) &=& -\frac{3}{2} \left( \text S - \text P \right) +\frac{3}{2}  \left( \text V_\parallel \right) - \frac{1}{2} \left( \text V _\perp \right)\,,
\ee
where
\be
 \left( \text S \CC - \text P \CC \right) &=& (\psib \CC \psib^T) (\psi^T \CC \psi)\, \nn\\
 && \qquad - (\psib \gamma_5 \CC \psib^T) (\psi^T \CC \gamma_5 \psi)\,,\\
 \left( \text A _\parallel \CC \right) &=& (\psib\gamma_0 \gamma_5 \CC \psib^T) (\psi^T \CC \gamma_0 \gamma_5\psi)\,,\\
 \left( \text A _\perp \CC \right) &=& (\psib\gamma_i \gamma_5 \CC \psib^T) (\psi^T \CC \gamma_i \gamma_5\psi)\,,
 \ee
with $\CC = \I \gamma_2 \gamma_0$ being the charge conjugation operator.

\section{RG Formalism}\label{app:RG}

For our computation of the RG flow equations 
of the various couplings and renormalization factors, 
we have employed the {\it Wetterich} equation~\cite{Wetterich:1992yh} which is an RG equation for the
(scale-dependent) quantum effective {action~$\Gamma_k$:
\be
\!\!\!\!\!\!\del_t \Gamma_k[\Phi] &=& -\frac{1}{2} \text{Tr} \left \lbrace [ \Gamma^{(1,1)}_k [\Phi] + R_k^{\psi} ] ^{-1}\cdot (\del_t R_k^{\psi}) \right \rbrace\,.
\label{eq:WetterichEquation}
\ee
Here,~$\Gamma^{(1,1)}_k$} denotes the second functional derivative of the (scale-dependent) quantum effective action~$\Gamma_k$
with respect to the fermion fields summarized in the ``super" {vector $\Phi^T(q) = \left(\psi^T (q),\bar{\psi}(-q)\right)$.}

\subsection{Regulator Function and Fermi Surface}\label{app:regfs}

The explicit calculation of RG flow equations requires the specification of the regulator function $R_k^{\psi}$ which encodes 
the regularization scheme. 
{In this work, we} have employed a four-dimensional {\it Fermi}-surface-adapted {regulator, specifically tailored
for a study of four-fermion interactions.} 
For our construction of this regulator
taking into account the presence of the {\it Fermi} surface, we start with an 
analysis of the spectrum of the fermionic kinetic term in the action {(including the chemical potential~$\mu$):
\be
\hat{T} = -( p\fslash + {\rm i}\gamma_0\mu)\,.
\ee
This} operator has four eigenvalues. The eigenvalues are partially degenerate. In fact, {there are only two} 
distinct pairs of {eigenvalues:
\be
\epsilon_{1,2}=\pm \sqrt{ (p_0+{\rm i}\mu)^2 +\vec{p}^{\,2}}\,.
\ee
For~$p_0=0$, we} note that the eigenvalues assume the following form:
\be
\epsilon_{1,2}\big|_{p_0=0}=\pm \sqrt{\vec{p}^{\,2}- \mu^2}\,.\label{eq:zeroevfree}
\ee
Thus, for~$p_0=0$, the eigenvalues tend to zero for momenta close to the {\it Fermi} momentum~$\mu$. Moreover, we note that
the eigenvalues are in general complex-valued quantities at finite~$\mu$.

We now construct a regulator function which also takes into account the presence of a potential zero mode 
at the {\it Fermi} surface (i.e. $p_0=0$), see Eq.~\eqref{eq:zeroevfree}.
To this end, we first note that  
the fermion propagator appearing in the loop integrals can be written in terms of the {eigenvalues~$\epsilon_{1,2}$:
\be
\!\!\!\!\!\!\!\!\!\!\frac{1}{p\fslash +\! {\rm i}\gamma_0\mu} =  \frac{p\fslash +\! {\rm i}\gamma_0\mu}{ \epsilon_{1,2}^2}
= \frac{(p\fslash +\! {\rm i}\gamma_0\mu)((p_0\! -\!\I\mu)^2 +\vec{p}^{\,2})}{\omega_{+}^2\omega_{-}^2}\,,
\ee
where}
\be
\omega_{\pm}^2\equiv \omega_{\pm}^2(p_0,\vec{p}^{\,})  = p_0^2 + ( |\vec{p}^{\,}| \pm \mu )^2\,.
\ee
Here,~$\omega_{\pm}$ is related to the quasiparticle dispersion relation associated with ungapped massless fermions:
For example,~$\omega_{-}(0,\vec{p})$ may be viewed as the energy required to create a particle 
with momentum~$\vec{p}$ above the {\it Fermi} 
surface. Correspondingly,~$\omega_{+}(0,\vec{p})$
is associated with the energy to create an antiparticle. Note that, for~$\mu=0$,~$\omega_{\pm}^2$ reduces to 
\be
\omega_{\pm}^2\big|_{\mu=0} = p_0^2 +\vec{p}^{\,2}\,.
\ee
For~$p_0\to 0$, we have
\be
\frac{1}{\epsilon_{1,2}^2} \bigg|_{p_0\to 0} \sim\, \frac{1}{\vec{p}^{\,2}-\mu^2}\,.
\ee
For our evaluation of the path integral with the aid of the {\it Wetterich} equation, 
we now construct a regularized kinetic {term:
\be
\hat{T}_{\rm kin} ^{\text{reg.}}= -( p\fslash + {\rm i}\gamma_0\mu)(1+r_{\psi})\,.
\ee
Here,~$r_{\psi}$ is} a so-called regulator shape function {and
\be
R^{\psi}_k= - (p\fslash + {\rm i}\gamma_0\mu) r_{\psi}\label{eq:Rcv1}
\ee
is} the associated regulator function. 

The regulator function is to a large extent at our disposal and 
only requires to fulfill a few constraints~\cite{Wetterich:1992yh}, see also below.
Assuming that~$r_{\psi}$ is 
a real-valued dimensionless function depending on~$p_0$, $\vec{p}$,~$\mu$, and the RG scale~$k$, {the regularized
eigenvalues are given by
\be
\epsilon_{1,2}^{\text{reg.}}= \pm \sqrt{ (p_0+{\rm i}\mu)^2 +\vec{p}^{\,2}}\, (1+r_{\psi})\,.
\ee
To} regularize the finite-$\mu$ zero modes appearing at any finite~$k$, see Eq.~\eqref{eq:zeroevfree}, we require that
\be
r_{\psi}\big|_{p_0=0\,, |\vec{p}^{\,}|\approx \mu} \sim \frac{1}{\sqrt{|\vec{p}^{\,2} -\mu^2|}}\,.\label{eq:regc1}
\ee
Moreover, we require that
\be
r_{\psi}\big|_{\mu=0,\{p_{\nu}\to 0\}} \sim \frac{1}{\sqrt{p_0^2+\vec{p}^{\,2}}}\,,\label{eq:regc2}
\ee
which ensures that the regulator function reduces to the conventionally employed covariant chirally symmetric
regulator functions in the limit~$\mu\to 0$. A specific choice for the shape function, which fulfills these 
conditions and has been employed in the present work, is given by
\be
r_{\psi} = \frac{1}{\sqrt{1-{\rm e}^{-\bar{\omega}_+ \bar{\omega}_-}}} - 1\,,\label{eq:Rcv2}
\ee
where $\bar{\omega}_{\pm} =\omega_{\pm}/k$.\footnote{In case of scale-dependent renormalization factors~$Z^{\parallel}$, $Z^{\perp}$, and~$Z_{\mu}$, the following
replacements in the definition of the regulator function~$R^{\psi}_k$ (including the shape function~$r_{\psi}$) may be required: $p_0\to Z^{\parallel}p_0$, $p_i\to Z^{\perp}p_i$,
and~$\mu \to Z_{\mu}\mu$.} We add that other shape functions, such as so-called {\it Litim}-type regulator 
functions~\cite{Litim:2000ci,Litim:2001fd,Litim:2001up}, can in principle be adapted accordingly by replacing~$p^2$ 
with~${\omega}_{+} {\omega}_{-}$.
In any case, with a regulator function fulfilling the constraints~\eqref{eq:regc1} and~\eqref{eq:regc2}, 
the eigenvalues of the kinetic term are finite at any finite value of~$k$.

Phenomenologically speaking, 
the so-defined class of shape functions also ensures that the momentum modes are integrated out around the {\it Fermi} surface,
similarly to regulator functions employed in RG studies of ultracold {\it Fermi} 
gases~\cite{Diehl:2007th} with spin- and mass-imbalance~\cite{Boettcher:2014tfa,Roscher:2015xha}.
This implies that modes with momenta~$|\vec{p}^{\,}|\simeq \mu$ are only taken into account in the limit~$k\to 0$ where
the regulator vanishes,~$R_k^{\psi}\to 0$. Thus, our regulator function screens 
modes with momenta close to the {\it Fermi} surface~$\mu$ but leaves modes with (spatial) 
momenta farther away from the {\it Fermi} surface unchanged. 
{We note that this class of shape functions also fulfills the standard requirements~\cite{Wetterich:1992yh}: 
\begin{itemize}
\item[(i)]
It remains finite in the limit of vanishing four-momenta.
\item[(ii)] 
It diverges suitably for~$k \to \infty$
to ensure that the quantum effective action approaches the classical action.
\item[(iii)]
It vanishes in the limit~$k\to 0$. 
\end{itemize}
In addition, our {\it Fermi}-surface-adapted 
class of regulator functions fulfills a set
of ``weak"/``convenience" requirements: 
\begin{itemize}
\item[(iv)] It does not violate the chiral symmetry of the 
kinetic term in the fermionic action.
\item[(v)] It does not introduce an artificial breaking of {\it Poincar\'{e}} invariance and, in particular, it preserves {\it Poincar\'{e}} invariance
in the limit~$T\to 0$ and~$\mu\to 0$.
\item[(vi)] It respects the invariance of relativistic theories under the transformation~$\mu \to -\mu$. 
\item[(vii)] It ensures that the regularization of the loop diagrams is local in terms of
temporal and spatial momenta at any finite value of the RG scale~$k$.
\end{itemize}
The requirement (vii) essentially corresponds}
to the fact that the regulator function defines the details of the {\it Wilsonian} momentum-shell integrations. 

Alternatively, one may have considered to use regulator functions which only act on the spatial 
momenta of the fermions~\cite{Braun:2003ii,Schaefer:2004en,Blaizot:2006rj,Litim:2006ag}. However,
this class of regulator functions introduces an artificial explicit breaking of {\it Poincar\'{e}} invariance in the RG flow even at zero temperature and chemical
potential, i.e. in the {\it Poincar\'{e}}-invariant limit, and therefore violates the requirement (v) above. The artificial 
explicit breaking of {\it Poincar\'{e}} invariance appears to be particularly severe as we {shall discuss in Sec.~\ref{app:covregvsspreg} below.} 
Moreover, this class of regulator functions lacks locality in 
the direction of temporal momenta and therefore violates the requirement (vii), i.e.
all temporal momenta are taken into account at any RG scale~$k$ whereas spatial momenta are restricted
to (small) momentum shells around the scale~$k\simeq |\vec{p}^{\,}|$. Fluctuation effects are therefore washed out
and the construction of meaningful expansion schemes of the effective action is complicated within such a regularization scheme.

\subsection{Silver-Blaze Property and Derivative Expansion}\label{sec:sbpdexp}

At finite chemical potential, regularization prescriptions face an additional complication, 
namely the so-called {\it Silver-Blaze} ``problem"~\cite{Cohen:2003kd}. This refers to
the fact that the free energy of, e.g., a fermionic system does not exhibit a dependence on the 
chemical potential at {\it zero} temperature, 
i.e. it remains as that of the vacuum, provided that the chemical potential is less than some critical value. The latter is determined
by the (pole) mass of the {lightest particle carrying a} finite charge associated with the chemical potential. This also so-called {\it Silver-Blaze} property 
of the free energy carries over to the {correlation functions (see, e.g., Ref.~\cite{Marko:2014hea}) and 
should in principle {also be preserved}
by the regulator functions employed
in RG studies~\cite{Fu:2015naa,Khan:2015puu}. Phenomenologically} speaking, this property 
simply states that the fermion density (corresponding to the difference in the numbers
of fermions and antifermions) remains zero at {\it zero} temperature while the chemical
potential is less than the fermion (pole) mass.
Mathematically speaking, the {\it Silver-Blaze} property is a consequence of the fact that fermionic theories
are invariant under the following transformation:\footnote{Here, we effectively treat the chemical potential as an external
constant background field.}
\be
\psib \mapsto  \psib \E^{-\I \alpha \tau}, \enspace \psi\mapsto \E^{ \I \alpha\tau} \psi\,, \enspace \mu \mapsto \mu + \I\alpha\,,
\label{eq:SBtrafo}
\ee
where~$\alpha$ parametrizes the transformation and~$\tau$
is the imaginary time. Setting~$\alpha=q_0$, Eq.~\eqref{eq:SBtrafo} 
immediately implies the following invariance of the partition sum~$\mathcal Z$:
\be
{\mathcal Z}\Big|_{\mu \to \mu + \I q_0} = {\mathcal Z}\,.\label{eq:sbztrafo}
\ee
Thus, the partition sum is invariant under a shift of the chemical potential~$\mu$
along the imaginary axis. Assuming that~$\mathcal Z$ is analytic, it follows 
that~$\mathcal Z$ does not depend on~$\mu$ at all. In particular, we deduce
from an analytic continuation
of Eq.~\eqref{eq:sbztrafo} from~$q_0$ to~$\I q_0$
that~$\mathcal Z$ does not depend on the actual value of the real-valued
chemical potential~$\mu$.

For the effective action~$\Gamma$, it follows in the same way from Eq.~\eqref{eq:SBtrafo} that
\be
\Gamma[ \bar{\psi} \E^{-\I q_0 \tau},\E^{ \I q_0\tau} \psi ]\Big|_{\mu \to \mu + \I q_0}  = \Gamma[\bar{\psi},\psi]\,,\label{eq:Grel}
\ee
and similarly for higher $n$-point functions as the latter are obtained from~$\Gamma$ by taking functional derivatives with respect to the fields
and setting them to zero subsequently. 

On the level of correlation functions, we recall that, for example, the two-point function has a pole at~$p_0^2=-m_{\rm f}^2$ at~$\vec{p}=0$, 
where~$m_{\rm f}$ is the (pole) mass
of the fermion. Thus, an analytic continuation of the two-point function in the complex $p_0$-plane 
is restricted to the domain~$|p_0|\leq m_{\rm f}$, as the pole mass is the singularity closest to the origin of the 
complex $p_0$-plane. From Eq.~\eqref{eq:Grel}, on the other hand, we find the following relation for
the two-point function:
\be
&&\Gamma^{(1,1)}(p_0-q_0,\vec{p};p_0^{\prime}-q_0,\vec{p}^{\,\prime})\big|_{\mu \to \mu+\I q_0}\nn\\
&& \qquad\qquad\qquad\qquad\quad = \Gamma^{(1,1)}(p_0,\vec{p};p_0^{\prime},\vec{p}^{\,\prime})\,,
\label{eq:g11rel}
\ee
where the four-momenta $(p_0^{(\prime)},\vec{p}^{\,(\prime)})$ are associated {with the ingoing and outgoing 
fermion lines, respectively.} Note that~$\Gamma^{(1,1)}$ is diagonal in 
momentum space, i.e.
\be
 &&\!\!\!\!\!\!\!\!\!\! \Gamma^{(1,1)}(p_0,\vec{p};p_0^{\prime},\vec{p}^{\,\prime})\nn\\
&&=\tilde{\Gamma}^{(1,1)}(p_0,\vec{p}^{\,})
(2\pi)\delta(p_0\!-\! p_0^{\prime})(2\pi)^3\delta^{(3)}(\vec{p}\!-\!\vec{p}^{\,\prime})\,.
\ee
For~$q_0=\I\mu$, Eq.~\eqref{eq:g11rel} implies
\be
&&\left(\lim_{\mu\to 0}\Gamma^{(1,1)}(p_0,\vec{p};p_0^{\prime},\vec{p}^{\,\prime})\right)\bigg|_{p_0^{\,(\prime)}\to p_0^{\,(\prime)}-\I\mu} \nn\\
&& \qquad\qquad\qquad\qquad\quad = \Gamma^{(1,1)}(p_0,\vec{p};p_0^{\prime},\vec{p}^{\,\prime})\,
\label{eq:sbcorrfct}
\ee
for~$\mu < m_{\rm f}$. The latter constraint follows from the 
definition of the pole mass:~$\Gamma^{(1,1)}=0$ for~$(p_0-\I\mu)^2 = -m_{\rm f}^2$, i.e. $p_0=\I(\mu\pm m_{\rm f})$, and~$\vec{p}=0$. 
Note that~$m_{\rm f}$ refers to the pole mass at~$\mu=0$. 
This line of argument can be generalized straightforwardly to higher $n$-point
functions. 

Overall, it follows that, at {\it zero} temperature and~$\mu < m_{\rm f}$, 
the free energy of the system does not exhibit a dependence on~$\mu$, 
as stated above~\cite{Cohen:2003kd}.
The~$\mu$-dependence of the correlation functions is trivially obtained by replacing the 
zeroth components of the four-momenta in the vacuum correlation functions
with suitably $\mu$-shifted zeroth components, see, e.g., Eq.~\eqref{eq:sbcorrfct}.
{This then implies} that these functions also do 
not exhibit a dependence on~$\mu$ due to the analytic properties of these functions for~$\mu < m_{\rm f}$.
An immediate consequence is that the renormalization of the fermion mass and the chemical potential are
directly related at~$T=0$ and~$\mu < m_{\rm f}$, see our discussion in Sec.~\ref{sec:model}.
In any case, these statements cannot be generalized to the finite-temperature case as the zeroth
component of the {\it Euclidean} four-momentum becomes discrete due to the compactification of the
{\it Euclidean} time direction and the analytic continuation entering the above line of arguments cannot be defined uniquely.

In this work, we {use an RG approach} to compute correlation functions. The details of the {\it Wilsonian} momentum-shell integrations
are determined by our choice for the infrared regulator~$R^{\psi}_k$, see Eq.~\eqref{eq:WetterichEquation}. 
The regulator induces a gap~$\sim k$ for the two-point function and renders the correlation functions $k$-dependent. 
Provided that $\mu < m_{\text{gap}}(k)$, 
the~$\mu$-dependence of the correlation functions at~$T=0$ is now trivially obtained by replacing~$p_0$ 
with~$(p_0 - \I\mu)$ in the vacuum correlation functions.
Here,~$m_{\text{gap}}$ denotes the $k$-dependent gap determined by the distance of the 
singularity closest to the origin in the complex $p_0$-plane at~$T=0$. For example, we have~$m_{\text{gap}}\sim k$
for massless fermions. For fermions with (pole) 
mass~$m_{\rm f}$, on the other hand, we have~$m_{\text{gap}} \to m_{\rm f}$ for~$k\to 0$ and~$m_{\text{gap}}\sim k$
for~$k\gg m_{\rm f}$. It follows that the RG flows of  
correlation functions for a given~$\mu < m_{\text{gap}}(k)$ at~$T=0$ 
are identical to their vacuum flows.

Our conclusions following from the invariance of the theory under the transformation~\eqref{eq:SBtrafo} are exact statements,
provided this invariance is not violated  
by the regularization scheme and the
expansion/approximation scheme. 

With respect to, e.g., the derivative expansion of the effective action, an expansion of the 
correlation functions
about the point~$(p_0-\I\mu,\vec{p}^{\,})=(0,0)$ rather than~$(p_0,\vec{p}^{\,})=(0,0)$ is required to 
preserve exactly the {\it Silver Blaze} property~\cite{Fu:2015naa}. This follows immediately from the fact that the 
correlation functions do not have an explicit $\mu$-dependence but depend only on the chemical potential 
via a $\mu$-shift of the zeroth component of the four momenta, see, e.g., Eq.~\eqref{eq:sbcorrfct}.
If the derivative expansion is nevertheless 
anchored at the point~$(p_0,\vec{p}^{\,})=(0,0)$, an explicit breaking of the invariance 
under the transformation~\eqref{eq:SBtrafo} is introduced which, however, has been found to be mild in  
RG studies of QCD low-energy models with conventional {spatial 
regulator functions~\cite{Fu:2015naa,Khan:2015puu,Strodthoff:2011tz,*Kamikado:2012bt,*Strodthoff:2013cua,*Fu:2016tey}. 
Note that the choice of the}
expansion point may be more delicate 
when conventional spatial regulator functions without an adaption due to the presence of a {\it Fermi} surface are employed, 
see, e.g., Refs.~\cite{Braun:2003ii,Schaefer:2004en,Blaizot:2006rj,Litim:2006ag} for a definition of this class of
{regularization schemes.} 
This class of regulators lacks locality in the direction of the zeroth component
of the four-momentum, i.e. the corresponding regulator functions are ``flat" in this direction and therefore
all time-like momentum modes effectively contribute to the RG flow at any value of~$k$. In fact, in our present analysis, we even observe
that the choice of the expansion point~$(p_0,\vec{p}^{\,})=(0,0)$
leads to ill-defined RG flows because of the analytic properties
of the threshold functions~$l^{\text{(F)}}_{\parallel \pm}$ and~$l^{\text{(F)}}_{\perp \pm}$ 
at~$T=0$ and~$\mu>0$, see App.~\ref{subsec:sr} and
the right panel of Fig.~\ref{fig:pmpp} for the Feynman diagram associated with these functions.  
The use of a suitably chosen expansion point, i.e. a point 
respecting the {\it Silver-Blaze} property, cures this problem~\cite{BraunLeonhPawlo}. 

Although a suitably chosen expansion point respecting the {\it Silver-Blaze} property may cure
such pathologies in case of spatial regulator functions, we should keep in mind 
a subtlety coming along with the choice of a particular expansion point. Usually,
we are interested in choosing a point for the expansion which is suitable to study a particular physical effect. 
This point may indeed be in conflict with the above considerations
regarding the {\it Silver-Blaze} property.
To be specific, we may only be interested in an evaluation of the fully momentum-dependent 
correlation functions for a specific
configuration of the external momenta. For an estimate of the phase structure of a given theory,
for example, the limit of vanishing external momenta may be considered for the two-point function in order to project
on screening masses.
This evaluation point may then serve as the anchor point for a derivative expansion but
violates the {\it Silver-Blaze} property as discussed above. On the other hand, the choice of an expansion
point respecting the {\it Silver-Blaze} property may require to include high orders in the derivative expansion
in order to be able to reach {reliably the actual point of physical interest which, as an expansion point, 
may violate the
{\it Silver-Blaze} property}. This is indeed the situation in many studies and it is also the case in our present 
work as we are interested in the evaluation of the four-fermion correlation functions in a specific limit
in order to estimate the phase structure. To be concise, we choose the limit of vanishing external momenta
as the expansion point. If we had chosen an expansion
point respecting the {\it Silver-Blaze} property, then we would have not been able to reach reliably our actual point of interest
at leading order of the derivative expansion. 
A detailed discussion of this issue is deferred to future work~\cite{BraunLeonhPawlo}. 

In cases where the full momentum dependence of the correlation functions is resolved,
the only source of an explicit breaking of the {\it Silver-Blaze} property is the regularization scheme.
With respect to the regularization scheme in RG studies, 
in addition to the requirements (i)-(vii) listed above, the eigenvalues of the (matrix-valued) 
regulator function~$R^{\psi}_k$ are required to be only functions 
of the spatial momenta~$\vec{p}$ and the complex variable $z=p_0-\I\mu$, $R^{\psi}_k = R^{\psi}_k(z,\vec{p}^{\,})$~\cite{Khan:2015puu}
in order to preserve the invariance of the theory under the transformation~\eqref{eq:SBtrafo}. 
This requirement is fulfilled by spatial regularization schemes (such as 
the class of regulator functions defined in 
Refs.~\cite{Braun:2003ii,Schaefer:2004en,Blaizot:2006rj,Litim:2006ag}) as these schemes simply do not
depend on~$z$ at all. 
However, our present class of four-dimensional/covariant {\it Fermi}-surface-adapted
regulator functions fulfilling the requirements (i)-(vii) listed in Sec.~\ref{app:regfs}
breaks explicitly the symmetry associated with the {transformation~\eqref{eq:SBtrafo} 
as it depends on~$\omega_{+}\omega_{-}=|(p_0-\I\mu)^2+\vec{p}^{\,2}|$.}
\begin{figure}[t]
\centering
\includegraphics[width=0.5\textwidth]{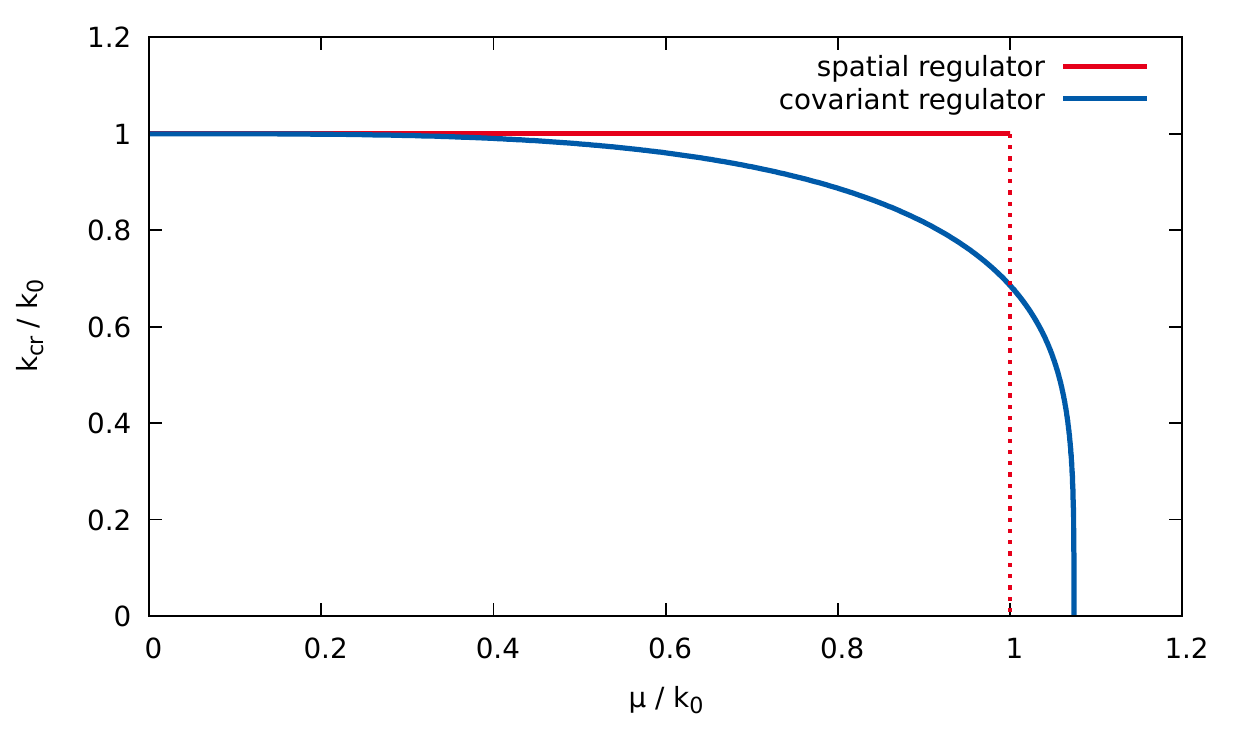}
\caption{Critical scale~$k_{\rm cr}/k_0$ with $k_0=k_{\rm cr}(\mu=0)\approx 0.35\Lambda$ as a function of~$\mu/k_0$ at zero temperature 
for two different regularization schemes, see {main text for details}.}
\label{fig:kcr}
\end{figure}

Our discussion with respect to the derivative expansion and the regularization
schemes calls for an analysis of the strength of the explicit
breaking of the {\it Silver-Blaze} property in our present study.
To this end, we consider the RG flow of the scalar-pseudoscalar coupling~$\lambda_{\sigma}$ in a
simple one-channel approximation and compute the dependence of the scale~$k_{\rm cr}$ 
on the chemical potential~$\mu$ using the covariant regulator function 
defined by Eq.~\eqref{eq:Rcv2} and the spatial 
{regularization scheme defined} in Refs.~\cite{Braun:2003ii,Schaefer:2004en,Blaizot:2006rj,Litim:2006ag}. 
Recall that the scale~$k_{\rm cr}$
is defined as the scale at which the four-fermion coupling~$\lambda_{\sigma}$ diverges. The scale dependence of the $\lambda_{\sigma}$-coupling 
is governed by the following flow {equation:
\be
\partial_t \lambda_{\sigma} &=& 2\lambda_{\sigma} - 48v_4 ( 
l^{\text{(F)}}_{\parallel +}(\tau,0,-\I\tilde{\mu}_{\tau})\nn\\
&& \qquad\qquad\qquad + l^{\text{(F)}}_{\perp +}(\tau,0,-\I\tilde{\mu}_{\tau})
 \lambda_{\sigma}^2\,,\label{eq:SB1c}
\ee
where} $\tau=T/k$ and~$\tilde{\mu}_{\tau}=\mu/(2\pi T)$. The 
definition of the threshold functions~$l^{\text{(F)}}_{\parallel +}$ and~$l^{\text{(F)}}_{\perp +}$ 
for the two regularization schemes can be found in App.~\ref{app:RGtf}. 
Compared to the flow equation~\eqref{eq:ocappt0m0}, we do not include 
the threshold functions $l^{\text{(F)}}_{\parallel \pm}$ and~$l^{\text{(F)}}_{\perp \pm}$ associated with the loop diagram depicted 
in the inset of the right panel of Fig.~\ref{fig:pmpp} in this analysis
since, for the spatial regularization scheme, these threshold functions
lead to ``spurious" divergences in the integration of 
the RG flow equations at~$T=0$ due to a second-order 
pole at~$k=\mu$. We refer to  
App.~\ref{subsec:sr} for explicit representations of these functions. 
This behavior is associated with the presence of a zero mode in the two-point function at~$k=\mu$,
see Eq.~\eqref{eq:zeroevfree}. 
Note that, for any even infinitesimally small finite temperature, these functions are well-behaved, i.e. no
``spurious" divergences in the integration of
the RG flow equations appear. 
Still, the contributions from these threshold functions
become arbitrarily large at~$\mu>0$ for decreasing temperature and therefore dominate artificially the 
RG flow of the couplings at finite chemical potential and low temperature.
Note that this is not the case for our covariant regulator function, which is well-defined for~$T=0$ and~$T>0$,
as it is constructed such that the zero mode at~$k=\mu$ is regularized.

Since $k_{\rm cr}(\mu)$ sets the scale for all low-energy observables including the fermion mass~$m_{\rm f}\sim k_{\rm cr}$ (see our 
discussion in Sec.~\ref{sec:fpsb}),
$k_{\rm cr}(\mu)$ should be independent of~$\mu$ for~$\mu < m_{\rm f}$ at {\it zero} temperature
because of the {\it Silver-Blaze} property. 
{Unfortunately, we do not have direct access to the fermion mass~$m_{\rm f}$ in our present study. However,
at least at zero temperature and chemical potential, the RG flow equation~\eqref{eq:SB1c} for the $\lambda_{\sigma}$ 
in the one-channel approximation can be mapped onto a corresponding mean-field equation for the fermion mass, 
see, e.g., Ref.~\cite{Braun:2011pp}. This provides us at least with an estimate for 
the vacuum fermion mass~$m_{\rm f}$ in our studies. Specifically, we find~$m_{\rm f}/k_0\approx 0.53$
for the covariant regulator function and~$m_{\rm f}/k_0\approx 0.44$ for the spatial regulator. Note that~$k_0$ has been fixed 
to the same value in both calculations.}
In Fig.~\ref{fig:kcr}, we show~$k_{\rm cr}$ as a function of~$\mu$ at zero 
temperature for the covariant regulator function
defined in Eq.~\eqref{eq:Rcv2} and the spatial 
{regularization scheme} defined in Refs.~\cite{Braun:2003ii,Schaefer:2004en,Blaizot:2006rj,Litim:2006ag}. In order to ensure
comparability, we have fixed the initial condition of the flow equation~\eqref{eq:SB1c} such that 
the symmetry breaking scale $k_0=k_{\rm cr}(\mu =0)$
assumes the same value in both cases. In accordance with our discussion, we observe that~$k_{\rm cr}$ 
does not depend on~$\mu$ for~$\mu < k_0$,
where~$k_0$ plays the role of the zero-temperature fermion mass.
Thus, this class of spatial regularization schemes in general respects the symmetry~\eqref{eq:SBtrafo} at zero temperature,
as already mentioned above. For our covariant
regulator function, we observe that~$k_{\rm cr}$ exhibits a weak {dependence on~$\mu$ for~$\mu \lesssim m_{\rm f}$. This}
dependence becomes stronger for increasing~$\mu$. For~$\mu/k_0\to 1$, $k_{\rm cr}$ then does not terminate but tends to zero continuously
at~$\mu/k_0\approx 1.1$. In any case, we find in both cases 
that the critical scale~$k_{\rm cr}$ is only finite for chemical potentials below some critical value~$\mu_{\rm cr}/k_0 \sim {\mathcal O}(1)$.

We emphasize that the artificial regulator-induced dependence on the chemical potential illustrated in Fig.~\ref{fig:kcr}
is an immediate consequence of the fact that our covariant regulator function violates the 
{\it Silver-Blaze} property. This violation becomes {evident by the fact that the} four-fermion
couplings depend on the chemical potential~$\mu$ at~$T=0$ for any value of~$k$. 
For~$k \gg \mu$, for example, we indeed deduce from Eq.~\eqref{eq:SB1c} that
\be
&& \lambda_{\sigma} \simeq \lambda_{\sigma}^{\ast} \bigg( 1
  + c_0\left(\frac{\mu}{k}\right)^2\ln\left(\frac{k}{\Lambda}\right)\nn\\
&& \qquad\qquad\; - \left(\frac{\Lambda}{k}\right)^2 \left(  \left(\frac{ \lambda_{\sigma}^{\ast}}{\lambda_{\sigma}^{(\rm UV)}} \right) -1 
\right)   
+\dots \bigg)\,,
\ee
where~$c_0<0$ is a numerical constant. 
 
In a calculation of the pressure at zero temperature, 
the violation of the {\it Silver-Blaze} property can also be observed. 
From a direct integration
of the flow equation~\eqref{eq:WetterichEquation} for the {\it free} gas, for example,
we obtain the correct result for the pressure~$p_{\rm free}$,
\be
p_{\rm free} = \frac{7}{2}\frac{\pi^2}{90}T^4 + \frac{1}{6}\mu^2 T^2 +\frac{1}{12\pi^2}\mu^4
\,,
\ee
if we choose the following initial condition for the effective action:
\be
\!\!\!\!\!\!\Gamma_{\Lambda}=-\frac{1}{2}\text{tr}_{\rm D}\sum_{n=-\infty}^{\infty}\int\frac{d^3p}{(2\pi)^3}\ln \left(
\Gamma^{(1,1)}_{\Lambda} + R_{\Lambda}^{\psi}\right)\,.
\ee
{with~$\Gamma^{(1,1)}_{\Lambda} =-( (\nu_n+\I\mu)\gamma_0 + \vec{p}\fslash)$ and~$\nu_n=(2n+1)\pi T$.}
The trace~$\text{tr}_{\rm D}$ has to be taken
with respect to the {\it Dirac} indices. In this case, the violation of the {\it Silver-Blaze} property  
becomes evident from the fact that~$\Gamma_{\Lambda}$ contains $\mu$-dependent divergent terms (e.g. $\sim\mu^2\Lambda^2$).
\begin{figure}[t]
\centering
\includegraphics[width=0.5\textwidth]{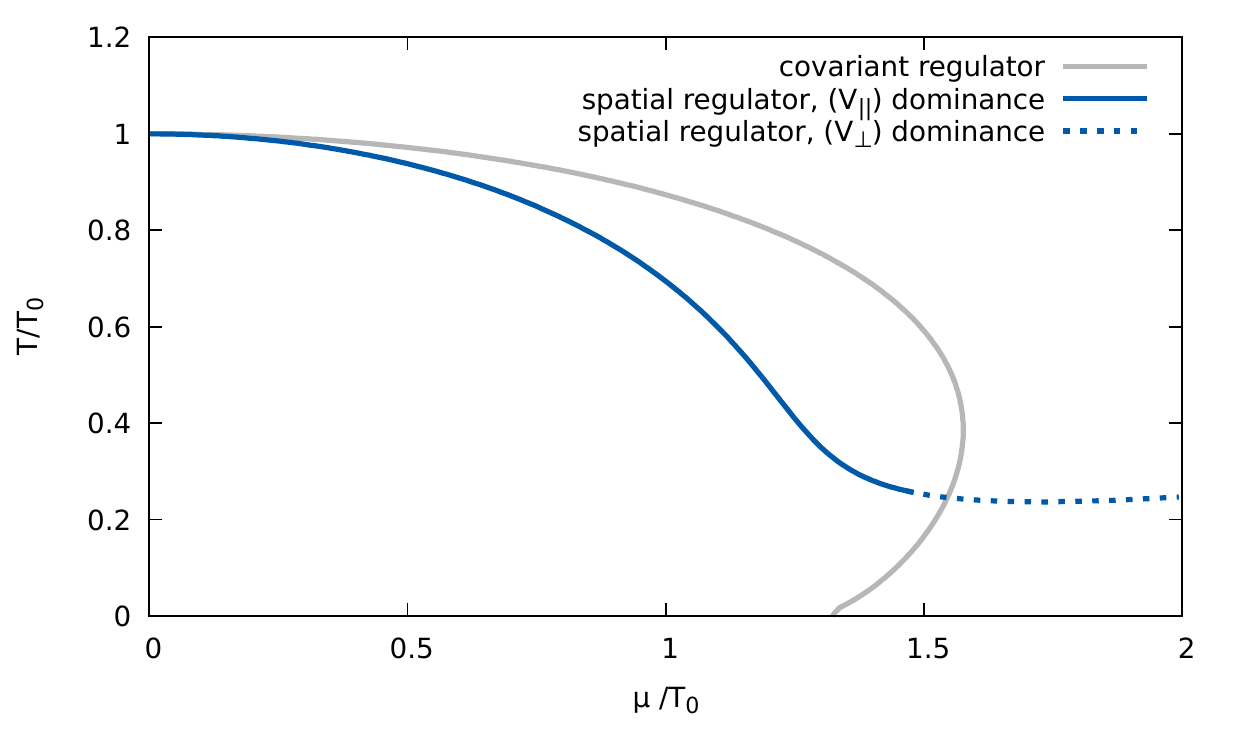}
\caption{Phase boundary associated with the spontaneous breakdown of at least one of the fundamental symmetries of our model 
as obtained from the {\it Fierz}-complete ansatz~\eqref{eq:FierzCompleteAnsatz} using 
two different regulator {functions, see Sec.~\ref{app:covregvsspreg} for a discussion
of the origin of the differences between the two phase boundaries. The gray line}
corresponds to the blue line in Fig.~\ref{fig:ComparisonChannels} and is only included to guide the eye.} 
\label{fig:pd3d}
\end{figure}
\subsection{Covariant Regulators versus Spatial Regulators}\label{app:covregvsspreg}

Our four-dimensional {\it Fermi}-surface-adapted 
regulator function defined by Eq.~\eqref{eq:Rcv2} violates the {\it Silver-Blaze} property.
Nevertheless, we have restricted 
ourselves to the use of this regulator
in our studies of the phase structure in Secs.~\ref{sec:fpsb} and~\ref{sec:ps}
since spatial regularization schemes
violate the requirements~(v) and (vii) listed in Sec.~\ref{app:regfs}, 
i.e. they introduce an explicit breaking of {\it Poincar\'{e}} invariance and they lack
locality in the direction of time-like momenta. 
Our four-dimensional {\it Fermi}-surface-adapted 
regulator fulfills both requirements.

In our {\it Fierz}-complete studies, 
we have indeed found that the artificial 
breaking of {\it Poincar\'{e}} invariance and the lack of locality 
affects the dynamics of the 
system already at zero chemical potential. {Even more, also} at~$T=\mu=0$, the~$\lambda_{\rm V}^{\parallel}$- 
and~$\lambda_{\rm V}^{\perp}$-coupling differ due to the explicit breaking of {\it Poincar\'{e}} invariance. 
This eventually results in a dominance of the~$\lambda_{\rm V}^{\parallel}$-channel at finite temperature and zero
chemical potential, see Fig.~\ref{fig:pd3d}. In our study with the 
covariant regulator function, on the other hand, we find a clear dominance of the $(\text{S}-\text{P})$-channel along the temperature axis
at~$\mu=0$. This aspect is of relevance
as such spatial regularization schemes may spoil the 
phenomenological interpretation of the results. {In fact,
at low temperature and large chemical potential, 
a study with the spatial regulator function even suggests that 
the dynamics of the system is strongly dominated by the $(V_{\perp})$-channel such that 
the ground state appears to be governed by spontaneous symmetry breaking for all values of the chemical potential considered
in this work ($\mu /T_0\lesssim 2$), see Fig.~\ref{fig:pd3d}. Using a different basis of four-fermion channels, e.g. including
difermion-type channels, one may even be tempted to associate the appearance of spontaneous symmetry breaking at (arbitrarily)
large chemical potential with the formation of a difermion condensate in our model
as it is the case in QCD (see, e.g., Refs.~\cite{Bailin:1983bm,Buballa:2003qv,Alford:2007xm,Anglani:2013gfu} for reviews).
In our present model, however, the appearance of a regime governed by spontaneous symmetry breaking
at large chemical potential is only observed when the spatial regulator function is used but not 
when our covariant {\it Fermi}-surface adapted regulator is applied. In fact, we consider the very appearance
of spontaneous symmetry breaking at large chemical potential in our present model as 
an artefact of the use of the spatial regulator function, at least at the order of the derivative expansion considered in this work.
Recall that the threshold functions 
$l^{\text{(F)}}_{\parallel \pm}$ and~$l^{\text{(F)}}_{\perp \pm}$ 
associated with the involved loop integrals are not well behaved at~$T=0$ and~$\mu>0$ in 
case of the spatial regulator, i.e. these threshold functions lead to ``spurious" divergences in the integration of the RG flow equations, 
see our discussion below Eq.~\eqref{eq:SB1c}. The definitions  
of the threshold functions in case of the spatial regulator are given in
 Sec.~\ref{subsec:sr}. From a phenomenological point of view, we 
note that, in contrast to QCD, the formation of a {\it Poincar\'{e}}-invariant
difermion condensate associated with $U_{\rm V}(1)$
symmetry breaking also entails chiral symmetry breaking in our present model, see Sec.~\ref{sec:ps}.}

{The relevance of covariant regularization schemes has also been discussed in the
context of real-time RG studies~\cite{Floerchinger:2011sc,Pawlowski:2015mia,Strodthoff:2016pxx}.
Along the 
lines of the construction of corresponding regulator functions~\cite{Pawlowski:2015mia},
it should in principle be
possible to construct a four-dimensional regulator} which fully respects the symmetry~\eqref{eq:SBtrafo} at zero temperature by
introducing a suitable deformation of the IR cutoff scale~$k$. However, this is beyond the scope of the present
work and deferred to future work~\cite{BraunLeonhPawlo}. 
Finally, we add that the complications
associated with the regularization of a theory in the presence of a finite chemical potential as well as the issues arising
because of the use of spatial regularization schemes are not bound to our functional RG approach but are in principle 
present in any approach.

\section{Threshold Functions}\label{app:RGtf}
In this section we define the threshold functions which appear in the RG flow equations in this work. 
The threshold functions essentially represent $1$PI diagrams and depend on the employed regularization scheme. In order to define these functions, 
we make use of various auxiliary dimensionless quantities, namely the dimensionless temperature $\tau = T/k$, the dimensionless (renormalized) 
chemical potential $\tilde \mu_{\tau} =  Z_{\mu}\mu / (2\pi T) $, and the dimensionless fermionic {\it Matsubara} frequencies $\tilde \nu _n = (2n+1) \pi \tau$. 

\subsection{Covariant Regulator}

It is convenient to define the dimensionless (regularized) propagator for the fermions:
\be
\tilde G_\psi (y_0,y,\omega) &=& \frac{1}{(y_0+y)(1+r_\psi)^2 + \omega}\,.
\ee
In the present work, the following purely fermionic threshold functions appear in the RG flow equations:
\begin{widetext}
\be
l^{\text{(F)}}_{\parallel +} (\tau, \omega, \tilde \mu_{\tau}) &=& -\frac{\tau}{2} \sum _{n=-\infty}^{+\infty} \int _0 ^{\infty} d y \, y^{\frac{1}{2}} \tilde \partial _t \left[  \left(\tilde \nu _n + 2 \pi \tau \tilde \mu_{\tau} \right)^2 (1+r_\psi)^2 \left( \tilde G _\psi ( (\tilde \nu _n + 2 \pi \tau \tilde \mu_{\tau} )^2 ,y,\omega)\right)^2 \right]\,,
\ee
\be
l^{\text{(F)}}_{\perp +} (\tau, \omega, \tilde \mu_{\tau}) &=& -\frac{\tau}{2} \sum _{n=-\infty}^{+\infty} \int _0 ^{\infty} \text{d} y \, y^{\frac{1}{2}} \tilde \partial _t \left[\left(y(1+r_\psi)^2 + \omega\right) \left( \tilde G _\psi ( (\tilde \nu _n + 2 \pi \tau \tilde \mu_{\tau} )^2 ,y,\omega)\right)^2 \right]\,, 
\ee
\be
l^{\text{(F)}}_{\parallel \pm} (\tau, \omega, \tilde \mu_{\tau}) &=& -\frac{\tau}{2} \sum _{n=-\infty}^{+\infty} \int _0 ^{\infty} \text{d} y  \, y^{\frac{1}{2}} \tilde \partial _t \bigg[ (\tilde \nu _n + 2 \pi \tau \tilde \mu_{\tau} )(\tilde \nu _n - 2 \pi \tau \tilde \mu_{\tau} )(1+r_\psi)^2 \nn \\ && \hspace{5.2cm} 
\times\, \tilde G _\psi ( (\tilde \nu _n + 2 \pi \tau \tilde \mu_{\tau} )^2 ,y,\omega)\tilde G _\psi ( (\tilde \nu _n - 2 \pi \tau \tilde \mu_{\tau} )^2 ,y,\omega) \bigg]\,,
\ee
\be
l^{\text{(F)}}_{\perp \pm} (\tau, \omega, \tilde \mu_{\tau}) &=& -\frac{\tau}{2}\!\sum _{n=-\infty}^{+\infty} \int _0 ^{\infty}\! \text{d} y \, y^{\frac{1}{2}} \tilde \partial _t \left[\left( y(1\!+\! r_\psi)^2 \!+\! \omega \right)\! \tilde G _\psi ( (\tilde \nu _n \!+\! 2 \pi \tau \tilde \mu_{\tau} )^2 ,y,\omega)\tilde G _\psi ( (\tilde \nu _n \!-\! 2 \pi \tau \tilde \mu_{\tau} )^2 ,y,\omega) \right]\,,
\ee
\end{widetext}
where $y = \vec{p}^{\,2} / k^2$ and
the formal derivative $\tilde \del _t$ is defined as~$\tilde \del _t = \left( \del_t r_{\psi}\right) \frac{\del}{\del r_{\psi}}$.
Here, we have already used that~$\partial_t Z^{\parallel} = \partial_t Z^{\perp} =\partial_t Z_{\mu} =0$ in our present study.
For the regulator function~\eqref{eq:Rcv2}, the latter assumes the following form:
\be
\tilde \del _t = \frac{(y_0+y) \E^{-(y_0+y)}}{(1\!-\!\E^{-(y_0+y)})^{\frac{3}{2}}}  \frac{\del}{\del r_\psi}.
\ee
In the limit~$\tilde{\mu}_{\tau}=0$, the above set of four distinct threshold functions collapses to a set of merely two threshold functions:
\be
&& l^{\text{(F)}}_{\parallel +} (\tau, \omega, 0) = l^{\text{(F)}}_{\parallel \pm} (\tau, \omega, 0 )  \equiv l^{\text{(F)}}_{\parallel} (\tau, \omega, 0)\,,\nn \\
&& l^{\text{(F)}}_{\perp +} (\tau, \omega, 0) = l^{\text{(F)}}_{\perp \pm} (\tau, \omega, 0)  \equiv l^{\text{(F)}}_{\perp} (\tau, \omega, 0)\,.\nn
\ee
{Furthermore, we find
\be
&&\!\!\!\!\!\!\!\!\! l^{\text{(F)}}_{\parallel} (\tau, \omega, 0) +  l^{\text{(F)}}_{\perp} (\tau, \omega, 0)\nn\\
&&\; = \tau \sum _{n=-\infty}^{+\infty} \int _0 ^{\infty} \text{d} y \, y^{\frac{1}{2}}  
\frac{(\del_t r_\psi) (1+r_\psi)( \tilde \nu_n ^2 + y) } { \left[ \left( \tilde \nu_n ^2 + y \right) (1 + r_\psi )^2 + \omega \right]^2}\,.
\ee
For} the regulator function~\eqref{eq:Rcv2} and~$T=\mu=\omega=0$, we then obtain~$l^{\text{(F)}}_{\parallel} (0, 0, 0) +  l^{\text{(F)}}_{\perp} (0, 0, 0)= \frac{1}{4}$.
In the limit~$T=\mu=\omega=0$, the threshold functions indeed only enter the RG flow equations in this particular combination.

\subsection{Spatial Regulator}\label{subsec:sr}
Also in case of  spatial regulator functions~\cite{Braun:2003ii,Schaefer:2004en,Blaizot:2006rj,Litim:2006ag},  
\be
R^{\psi}_k(p) =  -\vec{p}\fslash\, r_{\psi} \left({ \frac{\vec{p}^{\,2}}{k^2}}\right)\,
\label{eq:sregferm} 
\ee
with
\be
 r_{\psi}  =  \left(\sqrt{\frac{k^2}{\vec{p}^{\,2}}}\!-\!1\right)\theta(k^2\!-\! \vec{p}^{\,2})\,,\label{eq:3dsf}
\ee
it is convenient to define a {dimensionless propagator}:
\be
\tilde G^\text{spatial}_\psi (y_0,y,\omega) &=& \frac{1}{y_0+y(1+r_\psi)^2 + \omega}\,.
\ee
{The threshold functions then read}
\begin{widetext}
\be
l^\text{(F)}_{\parallel +,\text{spatial}} (\tau, \omega, \tilde \mu_{\tau}) &=& -\frac 1 2 \tau \sum _{n=-\infty}^{+\infty} \int _0 ^{\infty} d y \, y^{\frac{1}{2}} \tilde \partial _t \left[  \left(\tilde \nu _n + 2 \pi \tau \tilde \mu_{\tau}  \right)^2  \left( \tilde G^\text{spatial} _\psi ( (\tilde \nu _n + 2 \pi \tau \tilde \mu_{\tau} )^2 ,y,\omega)\right)^2 \right]\,,\label{eq:lsr1}
\ee
\be
l^\text{(F)}_{\perp +,\text{spatial}} (\tau, \omega, \tilde \mu_{\tau}) &=& -\frac 1 2 \tau \sum _{n=-\infty}^{+\infty} \int _0 ^{\infty} \text{d} y \, y^{\frac{1}{2}} \tilde \partial _t \left[\left(y(1+r_\psi)^2 + \omega\right) \left( \tilde G^\text{spatial} _\psi ( (\tilde \nu _n + 2 \pi \tau \tilde \mu_{\tau}  )^2 ,y,\omega)\right)^2 \right]\,,
\ee
\be
l^\text{(F)}_{\parallel \pm,\text{spatial}} (\tau, \omega, \tilde \mu_{\tau}) &=& -\frac 1 2 \tau \sum _{n=-\infty}^{+\infty} \int _0 ^{\infty} \text{d} y  \, y^{\frac{1}{2}} \tilde \partial _t \bigg[ (\tilde \nu _n + 2 \pi \tau \tilde \mu_{\tau}  )(\tilde \nu _n - 2 \pi \tau \tilde \mu_{\tau}  )\times \nn \\ && \hspace{3.2cm} \times\, \tilde G^\text{spatial} _\psi ( (\tilde \nu _n + 2 \pi \tau \tilde \mu_{\tau}  )^2 ,y,\omega)\tilde G^\text{spatial} _\psi ( (\tilde \nu _n - 2 \pi \tau \tilde \mu_{\tau}  )^2 ,y,\omega) \bigg]\,,
\ee
\be
\!\!\!l^\text{(F)}_{\perp \pm,\text{spatial}} (\tau, \omega, \tilde \mu_{\tau}) &=& -\frac 1 2 \tau \sum _{n=-\infty}^{+\infty} \int _0 ^{\infty} \text{d} y \, y^{\frac{1}{2}} \tilde \partial _t \left[\left( y(1+r_\psi)^2 + \omega \right)\times\right.\nn\\
&& \qquad\qquad\qquad\qquad\qquad
\left. \times\, \tilde G^\text{spatial} _\psi ( (\tilde \nu _n + 2 \pi \tau \tilde \mu_{\tau}  )^2 ,y,\omega)\tilde G^\text{spatial} _\psi ( (\tilde \nu _n - 2 \pi \tau \tilde \mu_{\tau}  )^2 ,y,\omega) \right]\,,\label{eq:lsr4}
\ee
\end{widetext}
{where $y = \vec{p}^{\,2} / k^2$ and}
\be
 \tilde \partial _t = \frac{1}{y^{\frac{1}{2}}}\theta(1-y)\frac{\partial}{\partial r_{\psi}}
\ee
for~$\partial_t Z^{\parallel} = \partial_t Z^{\perp} =\partial_t Z_{\mu}=0$. 
For the shape function~\eqref{eq:3dsf}, the threshold functions can be computed analytically.
For example, we find
\begin{widetext}
\be
l^\text{(F)}_{\parallel +,\text{spatial}} (\tau, \omega,  -\I \tilde{\mu}_{\tau})
&+& l^\text{(F)}_{\perp +,\text{spatial}} (\tau, \omega,  -\I \tilde{\mu}_{\tau}) \nn \\
&=& 
 \frac{1}{6}  \frac{\del }{\del \omega}\left\{
 \frac{1}{\sqrt{1+\omega}} \left[ \tanh \left( \frac{2\pi\tau\tilde{\mu}_{\tau} -\sqrt{1+\omega}}{2\tau} \right) - \tanh \left( \frac{2\pi\tau\tilde{\mu}_{\tau} + \sqrt{1+\omega}}{2\tau} \right)\right] \right\}\,, 
\ee
\be
&&\!\!\!\!\!\!\!\!\!\!\! l^\text{(F)}_{\parallel \pm,\text{spatial}} (\tau, \omega,  -\I \tilde{\mu}_{\tau})
+ l^\text{(F)}_{\perp \pm,\text{spatial}} (\tau, \omega,  -\I \tilde{\mu}_{\tau}) \nn \\
&& \quad =-\frac{1}{6}\frac{\partial}{\partial \omega}\left\{ \frac{1}{| \sqrt{1\!+\!\omega}-2\pi\tau\tilde{\mu}|}\tanh\left( \frac{| \sqrt{1\!+\!\omega}-2\pi\tau\tilde{\mu}|}{2\tau} \right)
+ \frac{1}{| \sqrt{1\!+\!\omega}+2\pi\tau\tilde{\mu}|}\tanh\left( \frac{| \sqrt{1\!+\!\omega}+2\pi\tau\tilde{\mu}|}{2\tau} \right) \right\}.
\label{eq:tf3dregpm}
\ee
\end{widetext}
Note that not only the sum of the two threshold functions in Eq.~\eqref{eq:tf3dregpm} has 
a second-order pole at~$\sqrt{k^2 + \omega}=\mu$ at~$T=0$ but also the individual functions. 
Moreover, in the limit~$\tau\to 0$, $\omega\to 0$, and~$\tilde{\mu}_{\tau}\to 0$, we find 
\be
\!\!\! &&l^\text{(F)}_{\parallel +,\text{spatial}} (0, 0, 0)
= l^\text{(F)}_{\perp +,\text{spatial}} (0, 0, 0) \label{eq:lsrlv} \\
&&\qquad = l^\text{(F)}_{\parallel \pm,\text{spatial}} (0, 0, 0)
= l^\text{(F)}_{\perp \pm,\text{spatial}} (0, 0, 0)
=\frac{1}{12}\,.\nn
\ee
\section{RG Flow Equations}\label{app:floweq}
For the derivation of the RG flow equations of our model, 
we have made use of existing software packages~\cite{Huber:2011qr,Cyrol:2016zqb}. 
For the covariant regulator function, we then find the following set of flow equations for the 
dimensionless four-fermion couplings $\lambda_\sigma$, $\lambda_\text V ^\parallel$ and $\lambda_\text V ^\perp$:
\begin{widetext}
\be
&&\partial_t \lambda _\sigma\equiv \beta _{\lambda _\sigma} = 2 \lambda_\sigma 
-16 v_4 
\left(-\lambda_\sigma ^2 + 2 \lambda_\text{V}^\parallel \lambda_\text{V}^\perp + (\lambda_\text{V}^\perp)^2 - 2\lambda_\sigma \lambda_\text{V}^\perp \right) 
l^{\text{(F)}}_{\perp \pm} (\tau, 0, -\I \tilde \mu_{\tau})
 \nn \\
&& \qquad\qquad\qquad\qquad  -16 v_4 
\left( 3\lambda_\sigma ^2+ 2 \lambda_\text{V}^\parallel (\lambda_\sigma +\lambda_\text{V}^\perp)+(\lambda_\text{V}^\perp)^2+ 8 \lambda_\sigma  \lambda_\text{V}^\perp \right)
l^{\text{(F)}}_{\perp +} (\tau, 0, -\I \tilde \mu_{\tau})
\nn \\
&& \qquad\qquad\qquad\qquad\qquad - 16 v_4 
\left(-\lambda_\sigma ^2 - 2 \lambda_\sigma  \lambda_\text{V}^\parallel + 3 (\lambda_\text{V}^\perp)^2\right)
l^{\text{(F)}}_{\parallel \pm} (\tau, 0, -\I \tilde \mu_{\tau})
\nn \\
&& \qquad\qquad\qquad\qquad\qquad\qquad -16 v_4 
\left( 3 \lambda_\sigma ^2 + 4 \lambda_\sigma  \lambda_\text{V}^\parallel + 3 (\lambda_\text{V}^\perp)^2 + 6 \lambda_\sigma \lambda_\text{V}^\perp \right)
l^{\text{(F)}}_{\parallel +} (\tau, 0, -\I \tilde \mu_{\tau} )\,, \label{eq:fefullls}
\ee
\be
&& \partial_t \lambda _\text{V}^\parallel\equiv \beta _{\lambda _\text{V}^\parallel} = 2 \lambda _\text{V}^\parallel 
+ 16 v_4 
\left(-\lambda_\sigma ^2 + 2 \lambda_\sigma  \lambda_\text{V}^\parallel + 4 \lambda_\text{V}^\parallel \lambda_\text{V}^\perp - (\lambda_\text{V}^\perp)^2 - 4 \lambda_\sigma  \lambda_\text{V}^\perp \right)
l^{\text{(F)}}_{\perp \pm} (\tau, 0,-\I \tilde \mu_{\tau} )
 \nn\\
&&  \qquad\qquad\qquad\qquad+ 16 v_4 
\left(-\lambda_\sigma ^ 2 - 2 (\lambda_\text{V}^\parallel)^2 - 2 \lambda_\sigma  \lambda_\text{V}^\parallel - 6 \lambda_\text{V}^\parallel \lambda_\text{V}^\perp - (\lambda_\text{V}^\perp)^2 - 4 \lambda_\sigma  \lambda_\text{V}^\perp \right)
l^{\text{(F)}}_{\perp +} (\tau, 0, -\I \tilde \mu_{\tau} )
\nn \\
&&  \qquad\qquad\qquad\qquad\qquad + 16 v_4 
\left(3\lambda_\sigma ^2 + (\lambda_\text{V}^\parallel)^2 + 6 (\lambda_\text{V}^\perp)^2 + 6 \lambda_\sigma  \lambda_\text{V}^\perp \right)
l^{\text{(F)}}_{\parallel \pm} (\tau, 0, -\I \tilde \mu_{\tau} )
\nn \\
&&  \qquad\qquad\qquad\qquad\qquad\qquad + 16 v_4
\left(-\lambda_\sigma ^2 + (\lambda_\text{V}^\parallel)^2 + 4 \lambda_\sigma  \lambda_\text{V}^\parallel + 6 \lambda_\text{V}^\parallel \lambda_\text{V}^\perp + 6 \lambda_\sigma  \lambda_\text{V}^\perp \right)
l^{\text{(F)}}_{\parallel +} (\tau, 0, -\I \tilde \mu_{\tau})\,,
\ee
{\be
&& \partial_t \lambda_{\text{V}}^\perp\equiv \beta _{\lambda_{\text{V}}^\perp} = 2 \lambda_{\text{V}}^{\perp} 
 -\frac{16}{3} v_4 
\left(-\lambda_\sigma ^2 - (\lambda_\text{V}^\parallel)^2 - 2 \lambda_\text{V}^\parallel (\lambda_\text{V}^\perp - \lambda_\sigma )- 10 (\lambda_\text{V}^\perp)^ 2 - 4 \lambda_\sigma \lambda_\text{V}^\perp \right)
l^{\text{(F)}}_{\perp \pm} (\tau, 0, -\I \tilde \mu_{\tau})
\nn \\
&&  \qquad\qquad\qquad\qquad -\frac{16}{3} v_4 
\left( 3 \lambda_\sigma ^2 + (\lambda_\text{V}^\parallel)^2 + 2 \lambda_\sigma  \lambda_\text{V}^\parallel + 10 (\lambda_\text{V}^\perp)^2 \right)
l^{\text{(F)}}_{\perp +} (\tau, 0,-\I \tilde \mu_{\tau})
\nn \\
&&  \qquad\qquad\qquad\qquad\qquad -16 v_4 
\left(\lambda_\sigma ^2 - 2 \lambda_\text{V}^\parallel \lambda_\text{V}^\perp - (\lambda_\text{V}^\perp)^2 + 2\lambda_\sigma  \lambda_\text{V}^\perp \right)
l^{\text{(F)}}_{\parallel \pm} (\tau, 0, -\I \tilde \mu_{\tau} )
\nn \\
&&  \qquad\qquad\qquad\qquad\qquad\qquad - 16 v_4 
\left( \lambda_\sigma ^2 + 4 \lambda_\text{V}^\parallel \lambda_\text{V}^\perp + 5 (\lambda_\text{V}^\perp)^2 + 6\lambda_\sigma  \lambda_\text{V}^\perp \right)
l^{\text{(F)}}_{\parallel +} (\tau, 0, -\I \tilde \mu_{\tau} )\,.\label{eq:fefulllvperp}
\ee}
\end{widetext}
In the limit of vanishing temperature and chemical potential, these RG flow equations simplify to 
\be
\beta _{\lambda _\sigma} &=& 2 \lambda_\sigma
- 4 v_4 \left( 2 \lambda_\sigma ^2 + 2 \lambda_\sigma ( \lambda_\text{V} ^\parallel + 3 \lambda_\text{V}^\perp) \right. \nn \\
&& \qquad\qquad\qquad\qquad \left.  + 3 (\lambda_\text{V}^\perp )^2 + 3 \lambda_\text{V} ^\parallel \lambda_\text{V}^\perp \right), 
 \label{eq:lst0f}  
\ee
\be
\beta _{\lambda _\text{V}^\parallel} &=& 2 \lambda _\text{V}^\parallel 
- 4 v_4 \left( \lambda_\sigma^2 + (\lambda_\text{V} ^\parallel ) ^2 \right. \nn\\
&& \qquad\qquad\quad
\left. + \lambda_\sigma ( - \lambda_\text{V}^\parallel + 3 \lambda_\text{V}^\perp )  \right), \label{eq:lvpt0f} 
\ee
\be
\beta _{\lambda_{\text{V}}^\perp} &=& 2 \lambda_{\text{V}}^{\perp} 
-4 v_4 \left( \lambda_\sigma^2 + (\lambda_\text{V}^\perp)^2 \right. \nn \\
&& \qquad\qquad\quad \left. 
+ \lambda_\sigma ( \lambda_\text{V}^\parallel + \lambda_\text{V}^\perp )  \right). \label{eq:lvtt0f} 
\ee
Choosing {\it Poincar\'e}-invariant initial conditions, i.e. choosing~$\lambda_{\text{V}}^{\parallel}=\lambda_{\text{V}}^{\perp}=\lambda_{\text{V}}$ 
at the initial RG scale, we deduce from Eqs.~\eqref{eq:fefullls}-\eqref{eq:fefulllvperp} that {\it Poincar\'e} invariance remains intact in
the RG flow:
\be
\beta_{\lambda_{\text{V}}} = \beta _{\lambda _\text{V}^\parallel} \Big|_{\lambda_\text{V}^\parallel = \lambda_\text{V}^\perp = \lambda_\text{V}} =
 \beta _{\lambda _\text{V}^\perp} \Big|_{\lambda_\text{V}^\parallel = \lambda_\text{V}^\perp = \lambda_\text{V}}\, \label{eq:crlimit2}
\ee
with
\be
\beta_{\lambda_{\text{V}}}=2 \lambda_\text{V}- 4 v_4 (\lambda_\sigma + \lambda_\text{V})^2\,.\label{eq:crlimit3}
\ee
The flow equations in case of the spatial regulator function~\eqref{eq:sregferm} are obtained from the flow
equations~\eqref{eq:fefullls}-\eqref{eq:fefulllvperp} by simply replacing the threshold 
functions 
with their
counterparts for the spatial regulator defined in Sec.~\ref{subsec:sr}.
However, note that Eqs.~\eqref{eq:lst0f}-\eqref{eq:crlimit3} are altered for the spatial regulator as
the actual values of the threshold functions for a given set of values of~$\tau$, $\omega$, and~$\tilde{\mu}_{\tau}$
depend in general on the details of the regularization scheme. For example, we find that
the values of the threshold functions associated with the two specific regulators used in this work
differ in the limit~$\tau\to 0$, $\omega\to 0$, and~$\tilde{\mu}_{\tau}\to 0$. 
In any case, we stress that Eq.~\eqref{eq:crlimit2} no longer holds for the spatial regulator function as the latter 
breaks explicitly {\it Poincar\'e} invariance, even at~$T=\mu=0$.

\bibliography{qcd}

\end{document}